\documentclass[11pt,a4paper]{article}

\usepackage[utf8]{inputenc}
\usepackage[T1]{fontenc}
\usepackage{amsmath,amssymb,amsthm}
\usepackage{graphicx}
\usepackage{float}
\usepackage{booktabs}
\usepackage{array}
\usepackage{natbib}
\usepackage[dvipsnames]{xcolor}
\usepackage[font=small,labelfont=bf]{caption}
\usepackage[hidelinks]{hyperref}
\usepackage{authblk}
\usepackage{geometry}
\usepackage{placeins}
\newcommand{\Teff}{T_{\mathrm{eff}}}
\newcommand{\Tcore}{T_{\mathrm{core}}}
\newcommand{\chisq}{\chi^{2}}
\newcommand{\dof}{\mathrm{dof}}
\newcommand{\kapp}{\kappa}
\newcommand{\demfile}{\texttt{quiet\_sun\_eis.dem}}
\newcommand{\demreg}{\texttt{demregpy}}

\title{The Quiet-Sun DEM Under Kappa:\\
Diagnostic Degeneracy and the Failure of the Conductive Closure}
\author[1,2]{Victor Edmonds}
\affil[1]{Final Stop Consulting LLC, Raleigh, USA\\\texttt{vedmonds@finalstopconsulting.com}}
\affil[2]{Ronin Institute for Independent Scholarship 2.0, Sacramento, USA}
\date{}

\begin{document}

\maketitle

\begin{abstract}
\noindent
For a plasma whose electron distribution carries a $\kapp \approx 2.5$ suprathermal tail, the Spitzer-H\"arm conductive closure does not exist: the conductive flux is the third velocity moment, carried by the tail, and the local conductivity integral diverges across the entire $\kapp \in [2,3]$ range --- the finite value the closed-form $\kapp$-conductivity returns at $\kapp = 2.5$ is an analytic continuation of that divergent integral, not a physical conductivity. \citet{edmonds2026a} places the quiet solar corona (QS) in this regime. The present paper takes that as its premise and establishes the two failures that follow for any plasma in the class: the standard EUV-DEM diagnostic cannot resolve such a plasma, and the conductive term of the standard QS energy budget has no valid form.

The diagnostic failure is demonstrated end-to-end. Three synthetic source families processed through the standard EUV imaging + regularized DEM inversion pipeline of \citet{hannah2012} --- a single-T $\kapp = 2.5$ probe, a multi-T $\kapp = 2.5$ source on the \citet{brooks2009} shape, and the Brooks-shape Maxwellian forward model --- recover FWHMs of 0.222, 0.305, and 0.319 in $\log T$ at $\chisq/\dof \leq 1.00$. The same pipeline applied to 80 real quiet-Sun AIA patches at solar minimum returns FWHM median 0.283 (range 0.230--0.401), stable across two solar-minimum dates to within 0.003 dex. All three synthetic sources sit inside or just below this distribution; the pipeline does not distinguish them. Independent of the FWHM degeneracy, two structural features emerge: an iron charge-state crossover at Fe XI ($\kapp$/Mxw $\approx 1$ across $\kapp \in [2, 3]$), and an EUV continuum reversal at AIA wavelengths --- suppressed near 94 \AA, enhanced at longer wavelengths --- contrary to the X-ray-extrapolated expectation of uniform $\kapp$ hardening.

The ionization-gated diagnostic structurally returns the effective temperature $\Teff$, the tail-weighted moment of the distribution. The Spitzer-H\"arm closure governs bulk transport and takes the bulk-core temperature $\Tcore = (\kapp - 3/2)/\kapp \cdot \Teff$ as physical input. The mismatch invites a temperature substitution that yields a concrete fluid-budget reduction --- mechanically correct and physically empty, because the coefficient it corrects has no convergent form --- a worked demonstration of the trap latent in any $\kapp$-corrected fluid budget, not a corrected number: it is the Fourier-law closure itself that fails, not merely its temperature input. Two QS-specific empirical pillars for impulsive heating --- DEM-width multi-thermality and the fluid-conductive-budget gap --- lose their structural assumptions, and the quiet-Sun budget question shifts to the non-local kinetic transport that no fluid closure represents.
\end{abstract}


\bigskip\hrule\bigskip

%==================================================================
\section{Introduction}\label{sec:intro}
%==================================================================

The differential emission measure of the quiet solar corona (DEM$(T)$: the squared electron density per unit log-temperature integrated along the line of sight; defined formally in Eq.~\ref{eq:forward}), recovered from EUV multi-channel imaging and spectroscopy, has a familiar shape. A broad peak near $\log T = 6.0$--$6.2$ ($\log_{10}$ of $T$ in kelvin throughout; the coronal range is $\log T \approx 5.5$--$7$; differences are quoted in dex, where 1 dex denotes one decade of base-10 log) dominates the distribution \citep{brooks2009,hannah2012,delzanna2018}, with a secondary hot component at $\log T \approx 6.7$--$6.8$ in off-limb QS analyses \citep{warrenbrooks2009}. Three inferences follow. The width of the main peak is read as evidence the corona is multi-thermal along every line of sight. The hot tail is read as evidence for distributed impulsive heating by unresolved nanoflares. The temperature structure feeds the Spitzer-H\"arm conductive-loss term in the standard \citet{withbroe1977} quiet-Sun energy budget, which has historically returned a gap between conductive losses and the Alfv\'en-wave flux measured at the coronal base. This inferential chain underwrites quantitative impulsive-heating constraints across multiple regimes; active-region nanoflare frequency-distribution work \citep{warren2012,cargill2014} applies the same methodology to AR DEMs, and the present paper engages the quiet-Sun branch of the same approach.

\begin{figure}[t]
\centering
\includegraphics[width=0.9\linewidth]{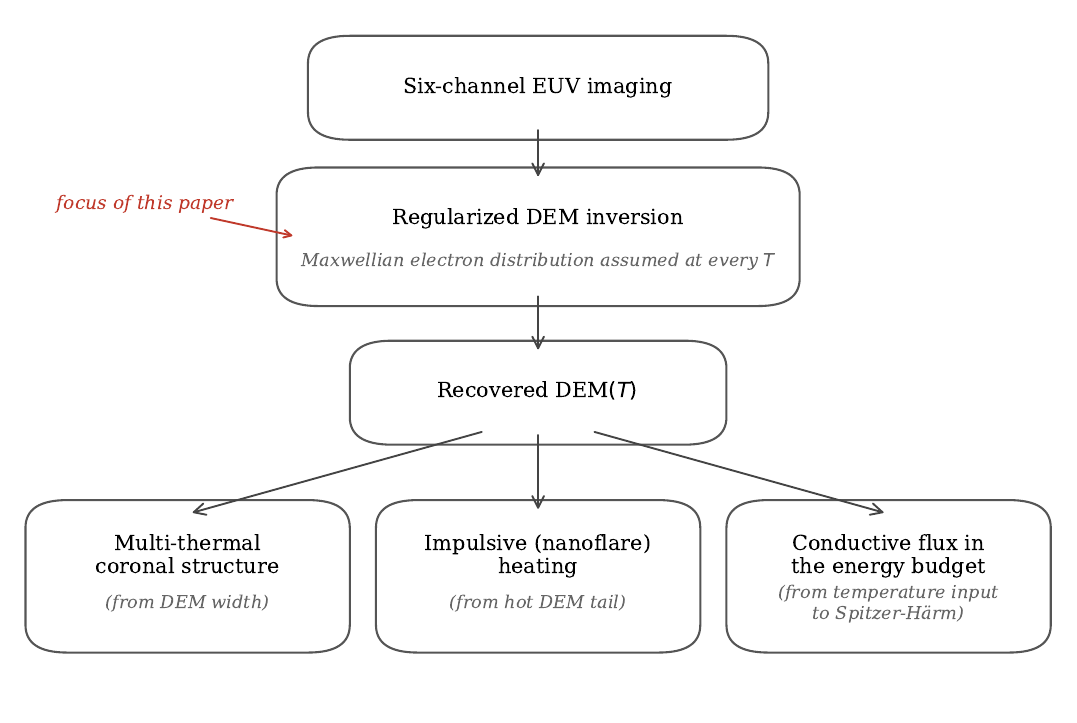}
\caption{The inference chain from EUV multi-channel imaging to coronal-heating constraints. SDO/AIA imaging produces channel photometry that is fed to the \citet{hannah2012} regularized DEM inversion. The inversion assumes a Maxwellian electron distribution at every temperature; the recovered $\mathrm{DEM}(T)$ is interpreted as the thermal structure of the emitting plasma, from which constraints on impulsive vs.\ steady heating are derived \citep{delzanna2018}, with analogous methodology underwriting active-region nanoflare frequency distributions \citep{warren2012,cargill2014}. The \citet{withbroe1977} standard quiet-Sun energy budget calculation uses the EUV-derived temperature in the Spitzer-H\"arm conductive-loss term. Each downstream inference inherits the Maxwellian assumption.}
\label{fig:inference}
\end{figure}

All three inferences pass through one computational step~(Fig.~\ref{fig:inference}): regularized inversion of channel photometry against a Maxwellian assumption at every temperature. \citet{edmonds2026a} provides empirical evidence that the quiet corona hosts a kappa-distributed suprathermal population (a power-law-tailed generalization of the Maxwellian, parameterized by $\kapp \geq 3/2$ controlling tail weight; $\kapp \to \infty$ recovers Maxwellian) with $\kapp \approx 2$--$3$, central value 2.5, derived from a three-diagnostic intersection of radio bremsstrahlung brightness temperatures ($T_{B} \approx 0.62$ MK at 150--450 MHz), EUV ionization-gated diagnostics ($\Teff \approx 1.5$ MK), and the hydrostatic scale-height density profile. Each diagnostic samples a different moment of the velocity distribution --- radio brightness probes the bulk, EUV ionization probes the tail, the hydrostatic profile probes the pressure-scale-height integral --- and each alone leaves $\kapp$ poorly constrained; the three-way intersection collapses the admissible region onto $\kapp \approx 2.5$ at the QS base. The Knudsen number in the quiet corona is $\mathrm{Kn} \equiv \lambda_{ee}/L_{T} \approx 0.01$--$0.1$ at typical QS conditions (electron mean free path $\lambda_{ee} \sim 10^{7}$ cm at $n_{e} \sim 10^{9}$ cm$^{-3}$ and $T \sim 1$ MK, against a temperature gradient scale length $L_{T} \sim 10^{8}$--$10^{9}$ cm); the upper edge is where Spitzer-H\"arm becomes unreliable and tail-electron mean free paths approach the gradient scale length \citep{shoub1983,vocks2016}. In the kinetic-theory sense the quiet corona is an extended Knudsen layer: the local fluid closure fails not only at the chromospheric interface but across the bulk, because the underlying distribution sits outside the family any local moment closure can accommodate (\S\ref{sec:closure-problem}). This paper assumes one thing and proves two. The assumption, from \citet{edmonds2026a} and defended there rather than here: the quiet corona is a $\kapp \approx 2.5$ plasma. The proofs, which hold for any plasma in that class: the standard DEM inversion is diagnostically degenerate against such a source (\S\ref{sec:results}), and the conductive closure behind the standard energy budget has no convergent form (\S\ref{sec:closure}).

\citet{edmonds2026b} establishes a convergence principle for ionization-gated electron-temperature diagnostics: any method whose inference is mediated by collisional ionization equilibrium structurally returns the effective temperature $\Teff$ of the underlying distribution and cannot detect departure from Maxwellian form. The mechanism is structural: collisional ionization requires electrons to clear an energy threshold $E_{\mathrm{thr}} \gg k\Tcore$, so the rate is exponentially gated by the suprathermal tail, and any diagnostic reading it is blind to the collisional bulk that governs fluid transport. Any distribution whose mean energy lands at a given $\Teff$ therefore produces the same charge-state distribution to within the slow shape-dependent variation of \citet{dudik2014b}'s within-ion ratios. DEM inversion, which reads channel photometry through ionization-set response functions, is one such diagnostic. The two results are established as follows. First (\S\ref{sec:results}), we generate synthetic SDO/AIA observations from a single isothermal $\kapp = 2.5$ source and run them through the standard \citet{hannah2012} regularized inversion to verify the convergence principle at the AIA-imaging level, identifying two features specific to the solar implementation. Second (\S\ref{sec:closure}), we show that the Spitzer-H\"arm conductive closure underlying the \citet{withbroe1977} quiet-Sun energy budget does not exist for a $\kapp \approx 2.5$ plasma, and demonstrate the temperature-substitution trap that the standard budget invites. Hinode/EIS spectroscopy with within-ion line ratios breaks the AIA-imaging degeneracy \citep{dudik2014b,dudik2019}; \citet{lorincik2020}'s QS EIS analysis is itself Maxwellian-consistent at the within-ion level --- the expected outcome under the convergence principle at the QS-relevant $\kapp$ range, and complementary to the present finding at the AIA-imaging level.

The scope is narrow. The Quiet Sun (QS: regions of the corona outside active regions, characterized by quasi-steady heating, low density $n_{e} \sim 10^{9}$ cm$^{-3}$, and low magnetic activity) is multi-thermal along every line of sight; the computational test in \S\ref{sec:results} uses a single isothermal $\kapp$ source as a probe of pipeline behavior, not as a model of the coronal plasma. The closure result of \S\ref{sec:closure} concerns fluid Spitzer-H\"arm conduction in the QS regime. Active regions (AR: magnetically intense regions with dynamic loops and higher density), where impulsive driving and dynamic loop structure dominate, are outside the scope. The total coronal energy budget the heating mechanism must supply depends on non-local kinetic transport that sits outside any fluid moment closure; quantifying it is the open kinetic-transport problem of \citet{edmonds2026a} \S 7 and is not addressed here.

%==================================================================
\FloatBarrier

\section{Method}\label{sec:method}
%==================================================================

The DEM pipeline is the standard solar-physics tool for inferring temperature structure from multi-channel EUV imaging. SDO/AIA \citep{lemen2012,boerner2012} observes the quiet corona in six EUV bandpasses centered at 94, 131, 171, 193, 211, and 335 \AA. Each channel has a temperature response function $K_{i}(T)$ (peaked sensitivity to plasma at a particular temperature with extended wings) derived from the atomic physics of the dominant ions in that bandpass under a Maxwellian electron distribution at temperature $T$. The forward problem is
\begin{equation}
\mathrm{DN}_{i} = \int K_{i}(T)\, \mathrm{DEM}(T)\, dT,
\label{eq:forward}
\end{equation}
where $\mathrm{DEM}(T) \equiv n_{e}^{2}(T)\, dh / d(\log T)$ is the differential emission measure: squared electron density along the line of sight per unit log temperature. Functionally, $\mathrm{DEM}(T)$ is the emission-weighted temperature distribution function along the line of sight, and Eq.~(\ref{eq:forward}) is an ill-posed inverse problem of recovering it from finite-channel integrals. Inversion solves for $\mathrm{DEM}(T)$ given the six $\mathrm{DN}_{i}$ (raw integrated photon counts at each AIA channel). The system is under-determined (six channels, $\sim$30 temperature bins); regularization picks the smoothest non-negative DEM consistent with the photometry. The output is a smooth $\mathrm{DEM}(T)$ curve interpreted as the thermal structure of the emitting plasma. Every step in this pipeline assumes the underlying electron distribution at each temperature is Maxwellian at that temperature; the question is what the pipeline returns when that assumption fails.

\subsection{Kappa ionization equilibrium}\label{sec:kappa-ioneq}

\begin{figure}[t]
\centering
\includegraphics[width=0.75\linewidth]{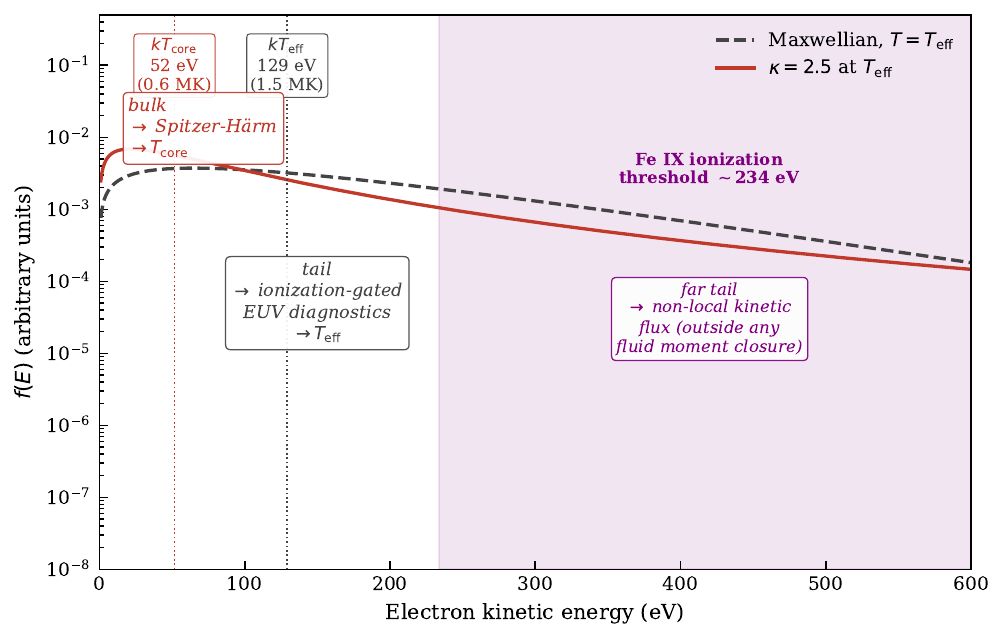}
\caption{The $\kapp = 2.5$ electron energy distribution (solid) and the Maxwellian at the same $\Teff$ (dashed), as functions of electron kinetic energy. The bulk core temperature $\Tcore = 0.6$ MK and the effective temperature $\Teff = 1.5$ MK are marked. Annotations show which part of the distribution each class of diagnostic samples: EUV ionization clears an energy threshold $E_{\mathrm{thr}} \gg k\Tcore$ and is gated by the suprathermal tail, returning $\Teff$; Spitzer-H\"arm fluid heat conduction weights low-velocity bulk integrands and takes $\Tcore$ as its physical temperature input; non-local kinetic flux carried by tail electrons sits outside any local fluid moment closure. The moment mismatch developed in \S\ref{sec:closure} is the consequence of feeding the wrong moment to the wrong closure.}
\label{fig:vdf}
\end{figure}

The $\kapp$ velocity distribution is the power-law-tailed generalization of the Maxwellian,
\begin{equation}
f_{\kapp}(\mathbf{v})
= n_{e}\, A_{\kapp}\, v_{th}^{-3}
\left[1 + \frac{v^{2}}{(\kapp - 3/2)\, v_{th}^{2}}\right]^{-(\kapp+1)},
\label{eq:kappa-vdf}
\end{equation}
with thermal speed $v_{th}^{2} \equiv 2 k_{B} \Teff/m_{e}$, normalization $A_{\kapp} = \pi^{-3/2}\, (\kapp - 3/2)^{-3/2}\, \Gamma(\kapp + 1)/\Gamma(\kapp - 1/2)$, and the Maxwellian recovered as $\kapp \to \infty$. The distribution requires $\kapp > 3/2$ for the mean energy to converge; the third velocity moment diverges at $\kapp \leq 2$ and the fourth at $\kapp \leq 5/2$ (\S\ref{sec:closure-nonexistence}). Figure~\ref{fig:vdf} plots the $\kapp = 2.5$ distribution against the Maxwellian at the same $\Teff$ and locates the diagnostic regimes the rest of the paper develops.

The $\kapp$ ionization equilibrium tables come from \citet{dzifcakova2013} and \citet{dzifcakova2023}, built on CHIANTI v10.1 atomic data \citep{dere2023}. The $\kapp = 2.5$ ion fractions are taken from the native Dz23 table (\texttt{Dz23\_kappa\_2p5.ioneq}). \citet{dzifcakova2023} state directly in their \S3.1: ``The value of $\kapp = 2.5$ was also added to bridge the relatively large gap between $\kapp = 2$ and 3.'' The table is computed from $\kapp$-distribution-integrated ionization and recombination rates with equilibrium solved directly. Not an interpolation on ion fractions.

\paragraph{Convention.} The \citet{dzifcakova2013} / \citet{dzifcakova2023} temperature parameter $T$ is the mean-energy temperature $\Teff$, defined by $\langle E \rangle = (3/2) k \Teff$. The core temperature of the $\kapp$ distribution is
\begin{equation}
\Tcore = \frac{\kapp - 3/2}{\kapp}\, \Teff.
\label{eq:Tcore}
\end{equation}
At $\log T = 6.176$ with $\kapp = 2.5$: $\Tcore = 0.6$ MK, $\Teff = 1.5$ MK. Every lookup in the Dz23 tables uses $\log T = 6.176$, not $\log T = 5.778$. Mixing these conventions flips the result. $\Teff$ is the primary parameterization throughout because it is the observational anchor: it is the moment that ionization-gated EUV diagnostics structurally return (\S\ref{sec:intro}), and the Dz23 tables themselves are parameterized by it. $\Tcore$ follows from Eq.~(\ref{eq:Tcore}) as a derived quantity; parameterizing by $\Tcore$ would place the observable downstream of a model parameter. The choice propagates into the DEM comparison explicitly: the thermal-speed parameter of Eq.~(\ref{eq:kappa-vdf}) is set by $\Teff$, so every $\kapp$-vs-Maxwellian comparison in this paper is made at matched observables --- $\kapp$-at-$\Teff$ against Maxwellian-at-$\Teff$ --- rather than at matched bulk temperature. As \S\ref{sec:closure} develops, $\Teff$ is also precisely the moment that bulk conduction does \emph{not} respond to; that mismatch between the measured moment and the transport-relevant moment is the second contribution of this paper.

The Dz23 EIMI and density-dependent DR suppression corrections are negligible at coronal densities ($n_{e} \approx 10^{9}$ cm$^{-3}$) and the v10.1 $\to$ v11 atomic-data shift is $<0.05\%$ for the ions that drive the analysis below.

\subsection{Per-ion synthetic emission}\label{sec:perion}

Each AIA channel under $\kapp$ is computed as a sum over ions:
\begin{equation}
\mathrm{DN}_{i}^{\kapp} = \sum_{\mathrm{ions}} \mathrm{DN}_{i,\mathrm{ion}}^{\mathrm{Mxw}}(\Teff) \cdot \frac{f_{\mathrm{ion}}^{\kapp}(\Teff)}{f_{\mathrm{ion}}^{\mathrm{Mxw}}(\Teff)},
\label{eq:perion}
\end{equation}
where $\mathrm{DN}_{i,\mathrm{ion}}^{\mathrm{Mxw}}$ is computed using ChiantiPy \citep{dere2023} with standard Maxwellian atomic physics at $\Teff$, and the ratio $f^{\kapp}/f^{\mathrm{Mxw}}$ is taken from matched Dz23 v10.1 ion-fraction tables.

\paragraph{Validity of the factorization approximation.} \citet{dudik2014b} computed $\kapp$-averaged collision strengths for strong coronal lines and found that within-ion strong-line ratios vary by a few tens of percent across $\kapp = 2$--25. The dominant $\kapp$ effect for AIA-channel-level analysis lives in the ion-fraction ratio $f^{\kapp}/f^{\mathrm{Mxw}}$ at $\Teff$, which captures $\sim$95\% of the channel response. The physical reason: line excitation samples energies $E \gtrsim E_{\mathrm{thr}}$ where $\kapp$-at-$\Teff$ and Maxwellian-at-$\Teff$ agree to $\lesssim 20\%$ \citep{dudik2014b}. The 20\% upper bound is the worst case across $\kapp \in [2, 25]$ and the full Fe IX--XIII strong-line set; at $\kapp = 2.5$ specifically and for the transitions dominating the bright AIA channels (Fe IX in 171 \AA, Fe XII in 193 \AA), where excitation energies lie near the formation threshold of the dominant ion at $\Teff$, the modification sits at the lower end of the cited range. The residual excitation-rate modification beyond the ion-fraction effect is bounded at this level; with the shot+read noise model of \S\ref{sec:noise} the per-channel uncertainty at low-DN channels (131 \AA{} at $\sim 5.6$ DN/s) is dominated by Poisson noise at $\sim 37\%$, well above the residual excitation-rate uncertainty.

A direct test is the Dz23 inversion of a true $\kapp = 2.5$ source, which returns 1.47 MK on a 1.5 MK input (2\% offset). The factorization is not appropriate for forbidden visible lines, where excitation-rate $\kapp$-modification is the leading effect (\S\ref{sec:openleft}). AIA channels also include hundreds of weak blends from non-Fe elements; their aggregate $\kapp$-modified excitation contribution is bounded by the same \citet{dudik2014b} tolerance.

\paragraph{Ion selection.} Inclusion criterion: ions with $\geq 0.1\%$ fractional abundance at $\Teff$ under $\kapp$, OR $\geq 0.1\%$ Maxwellian abundance anywhere in the AIA sensitivity range ($\log T = 5.5$--7.5). The pipeline computes ChiantiPy Maxwellian per-ion channel contributions for the broader admitted set; 165 unique ions produce a non-zero DN contribution in at least one AIA channel, yielding 984 (ion, channel) entries in the checkpoint.

\paragraph{Implementation.} Per-ion contributions are computed via ChiantiPy's \texttt{ion} class and folded through aiapy effective areas \citep{barnes2020}. The Maxwellian per-ion DN checkpoint is $\kapp$-independent; ion-fraction ratios are applied afterward.

\paragraph{Abundances.} Coronal abundances from the CHIANTI v11 \texttt{sun\_coronal\_2021\_chianti} compilation \citep{dere2023,dufresne2024}, which adopts photospheric values from \citet{asplund2021} with a 0.5-dex FIP-bias enhancement (the observed fractional enrichment of coronal abundances for elements with First Ionization Potential below $\sim 10$ eV relative to photospheric values) applied to low-FIP elements (Na, Mg, Al, Si, K, Ca, Sc, Ti, V, Cr, Mn, Fe, Co, Ni, Cu, Zn) following the standard \citet{feldman1992} prescription. Photospheric abundances \citep{asplund2021} are tested as a sensitivity check: $\Delta\chisq/\dof < 0.01$ (\S\ref{sec:sensitivity}).

\subsection{Continuum emission}\label{sec:continuum}

Free-free and free-bound continuum is computed with ChiantiPy's \texttt{continuum} class for 12 elements (H, He, C, N, O, Ne, Mg, Si, S, Ca, Fe, Ni) at $\Teff$ and folded through AIA effective areas. The continuum-to-line ratio is $\sim$5\% in 94 \AA, $\sim$23\% in 131 \AA, and $<1\%$ in 171, 193, 211, and 335 \AA. The 131 \AA{} fraction is large because of Fe recombination edges in the 90--140 \AA{} range.

The kappa/Maxwellian free-free ratio at photon energy $E$ is \citep{dudik2012}
\begin{equation}
\frac{j^{ff}_{\kapp}(E)}{j^{ff}_{\mathrm{Mxw}}(E)} = A_{\kapp}\, \left[1 + \frac{E}{(\kapp - 3/2)\, k\Teff}\right]^{-(\kapp+1)} \exp\!\left(\frac{E}{k\Teff}\right),
\label{eq:ffratio}
\end{equation}
with $A_{\kapp} = \Gamma(\kapp+1) / [\Gamma(\kapp - 1/2)\, (\kapp - 3/2)^{3/2}]$. The ratio is $<1$ when $E > k\Teff$ and $>1$ when $E < k\Teff$. At AIA wavelengths the $\kapp = 2.5$ values are 0.785 (94 \AA, $E = 132$ eV), 1.005 (131 \AA, 95 eV), 1.235 (171 \AA, 73 eV), 1.329 (193 \AA, 64 eV), 1.405 (211 \AA, 59 eV), and 1.802 (335 \AA, 37 eV). The crossover lies near 131 \AA{} where $E \approx k\Teff$. \S\ref{sec:euvreversal} returns to this energy-dependent behavior.

\paragraph{Free-bound treatment.} The free-bound continuum under $\kapp$ is computed per ion and per recombination edge by detailed balance. A free-bound photon of energy $E$ is emitted when an electron of kinetic energy $E_{e} = E - I_{\mathrm{edge}}$ is captured into a level of binding energy $I_{\mathrm{edge}}$; by the Milne relation the recombination cross-section is fixed by the photoionization cross-section and is independent of the electron velocity distribution. The entire $\kapp$-modification of the free-bound emissivity, per recombined level, is therefore the electron-flux ratio $f_{\kapp}(E_{e})/f_{\mathrm{Mxw}}(E_{e})$ at the \emph{electron} energy --- the functional form of Eq.~(\ref{eq:ffratio}) with argument $E_{e}$ rather than the photon energy. Near an edge $E_{e} \to 0$ and the ratio approaches $A_{\kapp} = 3.32$ at $\kapp = 2.5$, an enhancement of slow recombining electrons that a photon-energy scaling cannot capture; the 131 \AA{} band, sitting on the Fe VIII--X recombination edges, is where this matters most. The treatment is implemented as a per-level reweighting of the CHIANTI free-bound emissivities, applied as a ratio so per-ion normalization constants cancel, with two validation gates: the per-level decomposition reproduces the total Maxwellian emissivity shape exactly, and the in-band spectral ratio respects the analytic bound $A_{\kapp}$, reaching it only at edges. Evaluating the same machinery at the photon energy recovers the free-free scaling used as a bounding estimate in the submitted version, isolating the edge shift as the only change. Table~\ref{tab:freebound} reports the channel-integrated free-bound $\kapp$/Mxw ratios: at 131 \AA{} the edge-resolved ratio is 1.48 (ChiantiPy 0.15.2, the version of the submission baseline) to 1.58 (ChiantiPy 0.16.0), against 1.00 under the photon-energy proxy. Folding the recombining-ion population change through the Dz23 ion fractions gives $\approx 1.49$ on 0.15.2 and 3.64 on 0.16.0 under a reliability floor $f_{\mathrm{ion}}^{\mathrm{Mxw}} \geq 10^{-3}$; without the floor, bare and H-like trace ions with negligible Maxwellian presence and correspondingly unreliable equilibrium ratios (e.g.\ Mg XII at $f^{\mathrm{Mxw}} = 8 \times 10^{-6}$ with a Dz23 ratio of $3 \times 10^{3}$ and a recombination edge near 131 \AA) would dominate the channel; the guard mirrors the line-pipeline treatment of the same pathology.

The inversion-level impact is small everywhere: re-running the $\kapp = 2.5$ inversion with the per-ion free-bound in place of the proxy leaves the recovered peak at $\log T = 6.175$, the FWHM at 0.218, and $\chisq/\dof$ at 0.99--1.00 across the full computed range of treatments and atomic-code versions; the corrected 131 \AA{} DN (11.1--14.2 in pipeline units) sits inside the $3\times$ inflation envelope (16.9) tested in the submitted version, whose role is thereby reduced from proxy to bounding check. The pipeline absorbs the additional 131 \AA{} signal by lowering the per-channel recovery (96.9\% lines-only $\to$ 60.3\% with the per-ion treatment) rather than by reshaping the DEM.

\begin{table}[!htbp]
\centering
\caption{Channel-integrated free-bound $\kapp$/Mxw ratios at $\kapp = 2.5$, $\Teff = 1.5$ MK. ``Proxy'' is the submitted treatment (free-free ratio at the photon energy); ``per-ion'' is the edge-resolved detailed-balance treatment at the electron energy $E_{e} = E - I_{\mathrm{edge}}$, shown for the submission-baseline atomic code (ChiantiPy 0.15.2, canonical) and the current release (0.16.0); ``+ ion fractions'' folds the Dz23 recombining-ion population change under the $f_{\mathrm{ion}}^{\mathrm{Mxw}} \geq 10^{-3}$ reliability floor. All per-ion ratios are bounded by $A_{\kapp} = 3.32$ per edge. Exact reproduction of the canonical column requires ChiantiPy 0.15.2; the 0.16.0 column spans the cross-version band. The recovered DEM (peak, FWHM, $\chisq/\dof$) is unchanged across every column (see text).}
\label{tab:freebound}

\begin{tabular}{lrrrr}
\toprule
Channel & Proxy & Per-ion (0.15.2) & Per-ion (0.16.0) & + ion fractions (0.15.2 / 0.16.0) \\
\midrule
94 \AA  & 0.785 & 1.163 & 1.411 & 1.145 / 6.26 \\
131 \AA & 1.004 & 1.483 & 1.576 & 1.491 / 3.64 \\
171 \AA & 1.232 & 1.949 & 1.957 & 1.942 / 4.57 \\
193 \AA & 1.327 & 2.166 & 2.145 & 2.156 / 3.97 \\
211 \AA & 1.404 & 2.343 & 2.312 & 2.335 / 4.36 \\
335 \AA & 1.669 & 2.147 & 2.202 & 2.141 / 3.63 \\
\bottomrule
\end{tabular}

\end{table}

\subsection{AIA channel folding and noise model}\label{sec:noise}

Standard AIA temperature response functions are taken from \demreg{} (\texttt{aia\_tresp\_en.dat}, SSW/IDL-generated). Six coronal channels are used: 94, 131, 171, 193, 211, 335 \AA. Maxwellian response functions are used throughout; this replicates the standard analysis pipeline.

The noise model fed to \demreg{} as the per-channel uncertainty input combines two components: Poisson photon noise from the synthetic DN values at standard AIA exposure 2.9 s, and read noise of $[1.14, 1.18, 1.15, 1.20, 1.20, 1.18]$ DN (standard AIA per-channel values). Concretely, $\sigma_{i} = \sqrt{\sigma_{i,\mathrm{photon}}^{2} + \sigma_{i,\mathrm{read}}^{2}}$ at each channel. AIA absolute calibration and atomic-data uncertainty are quoted at $\sim 25\%$ per channel \citep{boerner2012,boerner2014}, but we do not add this as an independent systematic term to the \demreg{} input; the calibration uncertainty enters elsewhere (in the response functions themselves) and conflating it with the per-pixel statistical noise would inflate the uncertainty budget twice. Under shot+read alone, Poisson noise dominates at low-DN channels (e.g., 131 \AA{} fractional noise $\sim 37\%$ at $\mathrm{DN/s} \approx 5.6$) and read noise dominates only at the very brightest channels where photon counts are large. Per-channel residuals at the 30\%--40\% level (\S\ref{sec:deminversion}) sit within the natural shot-noise floor at low-DN channels and are not penalized by the global $\chisq$. With this noise model, $\chisq/\dof$ near unity is a real shot+read-noise-level fit, not the consequence of an inflated systematic budget. Adding the standard $\sim 25\%$ AIA systematic floor as an independent uncertainty would only loosen the fit further; shot+read is the tightest constraint, not the loosest.

\paragraph{Pipeline normalization vs.\ physical EM.} Synthetic DN magnitudes are generated at a pipeline normalization $\mathrm{EM}_{\mathrm{SCALE}} = n_{e}^{2} \cdot L = 10^{27}$ cm$^{-5}$, chosen to sit within the quiet-Sun white-light range. $\mathrm{EM}_{\mathrm{SCALE}}$ is a reference value, not a physical prediction. The physical EM that the $\kapp$ source actually requires to reproduce absolute line radiances is derived independently in \S\ref{sec:emrec} by reconciliation against \citet{brooks2009}, returning median $\log \mathrm{EM}_{\kapp} = 26.10$.

\subsection{DEM inversion}\label{sec:dem}

We use the \citet{hannah2012} regularized inversion via \demreg. Standard Maxwellian assumption throughout. Temperature grid $\log T = 5.7$ to $7.2$, step $0.05$ (31 bins). The inversion minimizes $\| K \cdot \mathrm{DEM} - \mathrm{DN}\|^{2} + \lambda \| L \cdot \mathrm{DEM} \|^{2}$ with $\lambda$ determined by generalized singular value decomposition.

The regularization strength is selected per generalized-singular-value mode by the Morozov discrepancy criterion (a target reduced $\chisq$), so no single scalar $\lambda$ characterizes an inversion; the operationally meaningful sensitivity test varies the discrepancy target. Re-running every synthetic source family at $0.5\times$, $1\times$, and $2\times$ the target changes the recovered FWHM by at most 0.009 dex (single-T $\kapp$: 0.006; multi-T $\kapp$: 0.003; Brooks-shape Maxwellian: 0.004; isothermal floor: 0.009 --- about one-fifth of a temperature bin) and does not move the recovered peak $\log T$ in any family. The recovered widths compared in \S\ref{sec:demshape} are not artifacts of the regularization-parameter choice.

Reported $\chisq/\dof$ divides the raw fit $\chisq$ by $(N_{\mathrm{channels}} - 1) = 5$. For a regularized inversion this is a convention: the actual effective number of degrees of freedom depends on the regularization strength and is not straightforwardly equal to any integer. The convention is used consistently across $\kapp$, abundance, and continuum-treatment comparisons so the reported values are internally comparable.

\subsection{Comparison target}\label{sec:comparison}

We compare against the \citet{brooks2009} Hinode/EIS quiet-Sun DEM, distributed as the CHIANTI reference model \demfile. The Brooks DEM has purely coronal coverage ($\log T = 5.4$--6.8) and is derived with coronal abundances and CHIANTI ionization equilibrium.

The absolute normalization of the recovered DEM depends on the assumed emission measure (arbitrary for the shape test). Both DEMs are normalized to their peak value and compared on shape metrics: peak $\log T$, FWHM in $\log T$, and high-temperature slope. Absolute-radiance consistency is addressed separately in \S\ref{sec:emrec}.

\paragraph{Real-QS reference sample provenance.} The matching-pipeline reference distribution of \S\ref{sec:demshape} is built from two solar-minimum dates. For each date (2019 December 1 and 2020 May 15, both near 12:00 TAI) one full-disk image per channel is taken from the \texttt{aia.lev1\_euv\_12s} series (nearest \texttt{QUALITY} $=0$ record), promoted to level 1.5 with standard \texttt{aiapy} registration (no network calibration corrections), and normalized to DN~s$^{-1}$ per pixel with per-channel exposure times ($\approx 2.9$ s; $\approx 2.0$ s for 171 and 193 \AA). Patches of $100'' \times 100''$ are placed on a deterministic $100''$ grid with centers within $0.95\,R_{\odot}$; patches with mean 171 \AA{} intensity above 500 DN~s$^{-1}$ (active region) or below 30 DN~s$^{-1}$ (coronal hole) are excluded, leaving 277 (2019) and 253 (2020) candidates, from which 80 evenly spaced patches per date are selected. No background subtraction is applied --- instrumental and scattered-light backgrounds are small relative to the on-disk quiet-Sun signal in these coronal channels, and a subtraction step would introduce a model dependence the comparison is designed to avoid; per-patch channel means feed the identical \demreg{} configuration and noise model used for the synthetic tests. The single-pixel shot+read model is applied to the patch mean by construction: it holds the real-data inversion to the same regularization constraint as the synthetic probe --- an apples-to-apples comparison of pipeline behavior on a typical quiet-Sun spectrum, not of the (much smaller) formal error of a $\sim\!10^{4}$-pixel average. Patch centers with per-patch results are archived with the analysis code. The recovered-FWHM distribution is stable across the two dates: medians 0.283 and 0.281 (shift $-0.0025$ dex), ranges 0.230--0.401 and 0.225--0.412, median peak $\log T = 5.975$ on both dates. The real-data inversions converge under the same shot+read noise model, with per-patch $\chisq/\dof$ median 0.998 (2019) and 1.003 (2020); fit quality was not a selection criterion --- no patch was excluded on $\chisq$.

%==================================================================
\FloatBarrier

\section{Results}\label{sec:results}
%==================================================================

This section is the empirical test of the convergence principle. By the convergence principle of \citet{edmonds2026b}, every diagnostic mediated by collisional ionization equilibrium returns $\Teff$ regardless of source family and is structurally degenerate in $\kapp$; DEM inversion is one such diagnostic, mapping EUV channel photometry through ionization-set response functions. The prediction is that the regularized AIA pipeline absorbs a single-T $\kapp$ source as apparent multi-thermal Maxwellian structure, with recovered FWHM inside the distribution the same pipeline returns from real QS observations. We verify the prediction (\S\ref{sec:fexi}--\S\ref{sec:demshape}), check absolute-radiance self-consistency against the published QS reference (\S\ref{sec:emrec}), and identify two features specific to the solar implementation: the iron charge-state crossover at Fe XI as a within-framework structural prediction (\S\ref{sec:fexi}), and an energy-dependent EUV free-free continuum reversal at AIA wavelengths reported in \S\ref{sec:euvreversal}.

%------------------------------------------------------------------
\subsection{Charge-state redistribution under kappa}\label{sec:fexi}
%------------------------------------------------------------------

\begin{figure}[t]
\centering
\includegraphics[width=0.85\linewidth]{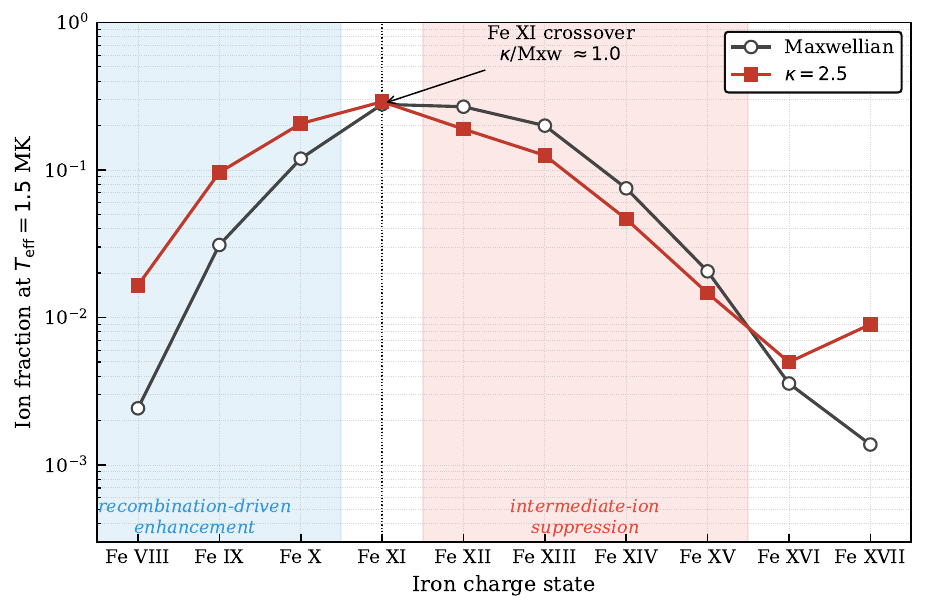}
\caption{Iron ion fractions versus charge state at $\Teff = 1.5$ MK. The Maxwellian distribution (open circles) and the $\kapp = 2.5$ distribution (filled squares) are overlaid on a logarithmic $y$-axis. The low-charge region (Fe VIII--X) is shaded as ``recombination-driven enhancement''; the high-charge region (Fe XII--XV) is shaded as ``intermediate-ion suppression.'' The Fe XI crossover ($\kapp$/Mxw $\approx 1.0$) is annotated.}
\label{fig:chargestates}
\end{figure}

Two competing processes set the iron charge-state distribution at any temperature (ion stages are denoted throughout by spectroscopic Roman numerals: Fe XII is the eleven-times-ionized iron atom, Fe IX is the eight-times-ionized iron atom, etc.). Ionization needs electrons above the relevant ionization threshold and is therefore driven by the suprathermal tail of the velocity distribution \citep{owocki1983}. Recombination has no threshold---any electron with finite kinetic energy can recombine---and is driven by the bulk \citep{hahnsavin2015}. Under a Maxwellian distribution, both processes sample the same exponential and arrive at the standard equilibrium. Under a kappa distribution at the same mean energy, the two processes sample different parts of the same distribution: the tail is enhanced relative to a Maxwellian at the same $\Teff$, and the bulk sits at lower energy than the Maxwellian peak. Ionization, hearing the tail, drives the equilibrium toward higher charge; recombination, hearing the cold bulk, drags it back. The two effects do not cancel: they redistribute the equilibrium into a characteristic U-shape across the iron ladder. Low-charge ions (Fe VIII--X at $\Teff = 1.5$ MK) are enhanced by bulk recombination, intermediate ions (Fe XII--XV) are suppressed where the two effects offset against the Maxwellian baseline, and very-high-charge ions (Fe XVI--XVII) are enhanced again by the tail driving ionization further than a Maxwellian-at-$\Teff$ would (Table~\ref{tab:feratios}, Fig.~\ref{fig:chargestates}). For iron at $\Teff = 1.5$ MK the lower crossover lands at Fe XI; a second crossover sits between Fe XV and Fe XVI where the tail-driven enhancement overtakes the intermediate-state suppression.

\begin{table}[!htbp]
\centering
\caption{Fe ion fraction ratios ($\kapp$/Mxw) at $\Teff = 1.5$ MK ($\log T = 6.176$), computed from the matched Dz23 v10.1 tables.}
\label{tab:feratios}
\begin{tabular}{lrrr}
\toprule
Ion & $\kapp = 2$ & $\kapp = 2.5$ & $\kapp = 3$ \\
\midrule
Fe VIII  & 21.4 & 6.8  & 3.8  \\
Fe IX    &  6.4 & 3.1  & 2.1  \\
Fe X     &  2.4 & 1.7  & 1.4  \\
Fe XI    &  0.95 & 1.04 & 1.00 \\
Fe XII   &  0.44 & 0.71 & 0.79 \\
Fe XIII  &  0.28 & 0.63 & 0.79 \\
Fe XIV   &  0.20 & 0.62 & 0.88 \\
Fe XV    &  0.16 & 0.71 & 1.12 \\
Fe XVI   &  0.26 & 1.40 & 2.31 \\
Fe XVII  &  1.14 & 6.52 & 9.83 \\
\bottomrule
\end{tabular}
\end{table}

\paragraph{Fe XI is structural.} The crossover at Fe XI is not a coincidence of $\kapp = 2.5$. The Fe XI ratio stays within 5\% of unity across $\kapp \in [2, 3]$ (0.954, 1.040, 0.998 for $\kapp = 2, 2.5, 3$), crossing unity once between $\kapp = 2$ and $\kapp = 2.5$. It is a structural feature: the ion at which tail-driven ionization and bulk-driven recombination balance. As a within-framework prediction, any single-$\kapp$ measurement at any $\Teff$ should show the charge-state crossover at the ion whose ionization potential places it at this balance point. The prediction is internal to the framework and observationally testable by independent atomic-physics measurements at other temperatures.

%------------------------------------------------------------------
\subsection{Synthetic AIA observations}\label{sec:dnratios}
%------------------------------------------------------------------

Each AIA channel has a dominant Fe ion. 171 \AA{} sees Fe IX. 131 \AA{} is dominated by Fe VIII at low charge with secondary contributions. 94 \AA{} sees Fe X. 193 \AA{} sees Fe XII. 211 \AA{} sees Fe XIII--XIV. 335 \AA{} sees Fe XIV--XVI in a configuration that straddles the Fe XI crossover; this channel is a known multi-ion blend in standard AIA processing, and the per-ion synthesis (Eq.~\ref{eq:perion}) includes all contributing ions explicitly so the channel-level results below incorporate the full blending pattern. The channel response under $\kapp$ therefore tracks the ion redistribution: channels whose dominant ions sit below the Fe XI crossover brighten; channels above the crossover dim; 335 \AA, straddling, barely changes.

\begin{table}[!htbp]
\centering
\caption{Per-channel $\kapp$/Mxw DN ratios (lines + continuum) for $\kapp = 2, 2.5, 3$.}
\label{tab:dnratios}
\begin{tabular}{lrrrl}
\toprule
Channel & $\kapp = 2$ & $\kapp = 2.5$ & $\kapp = 3$ & Dominant ion physics \\
\midrule
171 \AA  & 5.19 & 2.66 & 1.88 & Fe IX enhancement       \\
131 \AA  & 3.63 & 1.77 & 1.34 & Fe VIII enhancement     \\
94  \AA  & 1.87 & 1.37 & 1.18 & Fe X enhancement        \\
335 \AA  & 1.00 & 0.95 & 0.95 & Near crossover          \\
193 \AA  & 0.70 & 0.85 & 0.88 & Fe XII suppression      \\
211 \AA  & 0.48 & 0.74 & 0.86 & Fe XIII/XIV suppression \\
\bottomrule
\end{tabular}
\end{table}

\begin{figure}[t]
\centering
\includegraphics[width=0.85\linewidth]{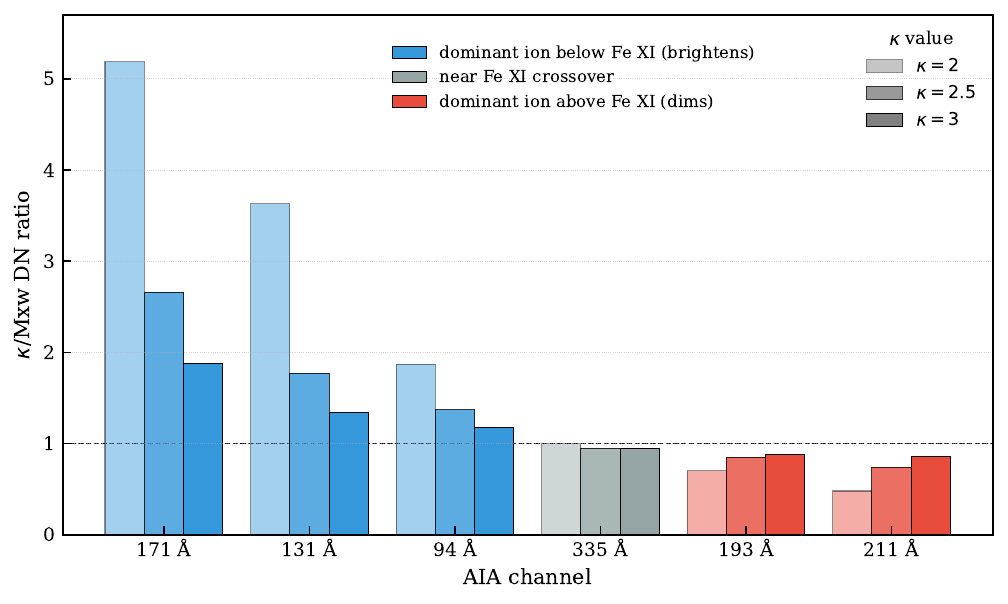}
\caption{Grouped bar chart of $\kapp$/Mxw DN ratios across the six AIA channels for $\kapp = 2, 2.5, 3$. Channels are colored by their dominant ion's position relative to the Fe XI crossover: 171, 131, 94 \AA{} (below crossover, brightening); 335 \AA{} (near crossover); 193, 211 \AA{} (above crossover, dimming). The visual pattern is the physical pattern of Table~\ref{tab:feratios}.}
\label{fig:channelratios}
\end{figure}

171 \AA{} is the most sensitive channel because Fe IX sits well below the Fe XI crossover~(Fig.~\ref{fig:channelratios}), where the ion-fraction ratio climbs steepest in $\kapp$. 211 \AA{} is the most strongly suppressed because Fe XIII--XIV sit well above the crossover, where the ratio drops most steeply. The 171/193 ratio compounds enhancement and suppression: $2.66 / 0.85 \approx 3.1\times$ shift relative to the Maxwellian baseline at $\kapp = 2.5$. This is the most $\kapp$-sensitive AIA-imaging observable identified in the analysis and is observationally accessible from existing data without new instrumentation (\S\ref{sec:openleft}). The $3.1\times$ shift applies to the strictly isothermal probe; under multi-thermal integration it largely averages out, because each channel samples its dominant ion near formation temperature where the ion-fraction ratio is close to unity (\S\ref{sec:demshape}). Consistently, the patch-mean 171/193 ratios measured in the real quiet-Sun sample of \S\ref{sec:comparison} (median 2.10, archived per patch) lie in the multi-thermal regime, far from the single-T probe value. A caveat for 193 \AA{} observers: the 193 channel includes a small transition-region (TR) plasma contribution from O V 192.90 \AA{} and other secondary blends. For quiet-Sun lines of sight these contributions are subdominant relative to Fe XII (the per-ion checkpoint shows them at $\lesssim 5\%$ of the total channel response at $\Teff$), so the 171/193 $\kapp$-shift remains a clean diagnostic in QS; it would be less reliable in regions with elevated TR plasma along the line of sight.

%------------------------------------------------------------------
\subsection{DEM inversion results}\label{sec:deminversion}
%------------------------------------------------------------------

The DEM inversion has thirty-one temperature bins to fit six channel measurements. It is heavily under-determined. Regularization picks the smoothest non-negative DEM consistent with the photometry. The $\kapp$ channel redistribution is itself smooth in temperature: it shifts the apparent thermal weighting toward lower-charge ions and away from higher-charge ones, but the shift is gradual along the temperature axis. The inversion has plenty of flexibility to absorb that shift by redistributing DEM weight across temperature bins. We should expect the $\chisq$ to come out near unity and the recovered DEM to be broader than a delta function, with a hot tail wherever the $\kapp$ excitation tail wants it.

\begin{table}[!htbp]
\centering
\caption{DEM fit quality across $\kapp$.}
\label{tab:fitquality}
\begin{tabular}{lrrrr}
\toprule
$\kapp$ & $\Tcore$ (MK) & $\chisq/\dof$ (lines) & $\chisq/\dof$ (lines + continuum) & DN recovery range \\
\midrule
3   & 0.75 & 1.02 & ---           & 87--112\% \\
2.5 & 0.60 & 1.52 & 1.00 & 63--110\% \\
2   & 0.38 & 3.49 & ---           & 61--113\% \\
\bottomrule
\end{tabular}
\end{table}

The assumption-light statement comes first: the recovered shape metrics agree between the lines-only and lines+continuum treatments to within 0.004 dex in FWHM at identical peak $\log T$ --- a shape-level agreement that does not inherit the noise normalization behind any single $\chisq$ value. At $\kapp = 2.5$ with continuum, $\chisq/\dof = 1.00$. The inversion finds a statistically acceptable Maxwellian DEM for synthetic data generated from a non-Maxwellian source. Both inversions are reported as alternates throughout: lines-only $\chisq/\dof = 1.52$ and lines+continuum $\chisq/\dof = 1.00$. The continuum-included value is the physical case --- continuum is real AIA signal (23\% of the 131 \AA{} channel), and with the per-ion free-bound treatment of \S\ref{sec:continuum} it is no longer a bounding estimate; lines-only is the conservative special case. The degeneracy conclusion holds in both treatments. At $\kapp = 3$, $\chisq/\dof = 1.02$---equally invisible. At $\kapp = 2$, $\chisq/\dof = 3.49$ (lines only) begins to strain the fit, though within what is routinely accepted given AIA calibration systematics.

\begin{figure}[t]
\centering
\includegraphics[width=0.7\linewidth]{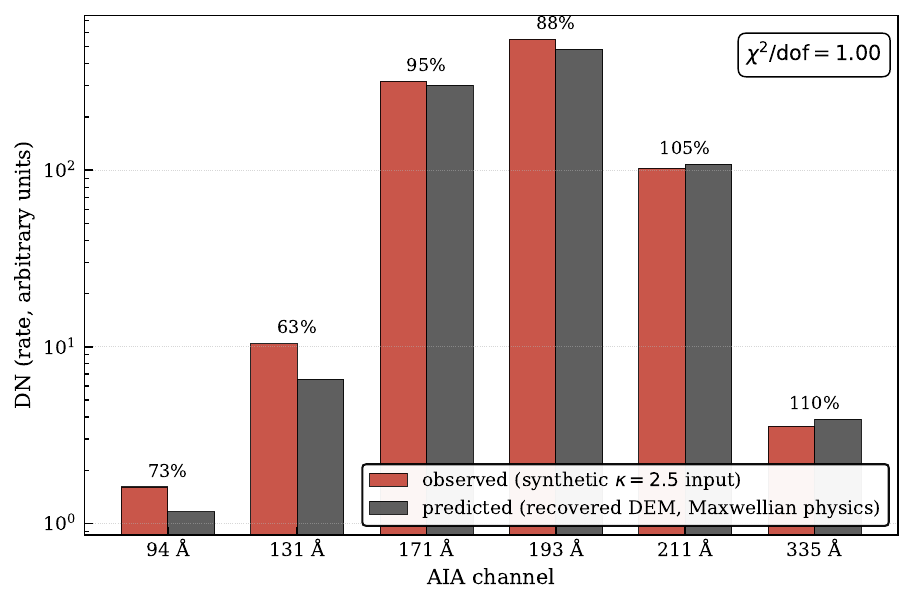}
\caption{Per-channel DN recovery for $\kapp = 2.5$: observed synthetic input versus prediction from the recovered DEM under Maxwellian physics, as paired bars across all six AIA channels. The pipeline's Maxwellian-predicted DN match the $\kapp$-generated input to within $\chisq/\dof = 1.00$, with per-channel recovery in the 63\%--110\% range (the 131 \AA{} recovery becomes 60\% under the per-ion free-bound treatment of \S\ref{sec:continuum}; \S\ref{sec:deminversion}). The recovered DEM shape itself --- the multi-thermal-looking curve that produces these matched DN --- is shown overlaid on the \citet{brooks2009} reference in Fig.~\ref{fig:headline}.}
\label{fig:deminversion}
\end{figure}

\paragraph{Where the $\kapp$ signature is hardest to absorb.} The 131 \AA{} channel shows the largest per-channel deviation~(Fig.~\ref{fig:deminversion}): 63\% DN recovery with continuum at $\kapp = 2.5$. This is where Fe VIII line excess (6.8$\times$) and the 23\% continuum fraction compound. The pipeline weights channels by their uncertainty, and 131 \AA{} is faint with large fractional noise; the global $\chisq/\dof$ remains near unity despite this per-channel deviation. With the per-ion free-bound treatment of \S\ref{sec:continuum} in place, the 131 \AA{} recovery moves from 63\% to 60\% and the global $\chisq/\dof$ from 1.00 to 0.99: the fully resolved treatment confirms rather than strains the fit.

%------------------------------------------------------------------
\subsection{Recovered DEM shape}\label{sec:demshape}
%------------------------------------------------------------------

Kappa charge-state broadening produces apparent thermal width because the inversion can only move emission along the temperature axis. When channels redistribute (171 brightens, 211 dims), the only structure available to the inversion to accommodate that pattern is to widen the DEM in temperature. The widening is two-ended: enhancement of the channels dominated by ions below the Fe XI crossover (Fe VIII--IX in 131 and 171 \AA) forces DEM weight toward low temperatures, enhancement of the tail-driven high ions (Fe XVI--XVII, Table~\ref{tab:feratios}) forces the hot component, and suppression of the intermediate ions hollows the middle --- the mechanism behind both the recovered width and the bimodal structure noted below. The amount of broadening should track the strength of the channel redistribution, which tracks the $\kapp$ value. We should expect $\mathrm{FWHM}(\kapp = 3) < \mathrm{FWHM}(\kapp = 2.5) < \mathrm{FWHM}(\kapp = 2)$, and the $\kapp = 2.5$ case should be the one whose broadening matches what observers measure when applying the AIA-imaging pipeline to the quiet Sun.

\begin{table}[!htbp]
\centering
\caption{DEM shape metrics, sorted by recovered FWHM. The isothermal-Maxwellian row is the algorithmic floor of the \demreg{} pipeline: a synthetic delta-function DEM at $\log T = 6.176$ recovered through the same response functions and regularization gives FWHM = 0.174, the irreducible width contributed by the AIA response kernels plus the GSVD-selected smoothing. All entries below it carry physical broadening above this floor. The floor-corrected column subtracts the algorithmic floor in quadrature, $\sqrt{\mathrm{FWHM}^{2} - 0.174^{2}}$ (the standard approximation for near-Gaussian broadening, conservative here), and applies only to pipeline outputs (``---'' marks the floor itself and the non-pipeline Brooks input shape). The same correction applied to the real-QS reference distribution gives median 0.223, range 0.150--0.361; the family ordering and mutual indistinguishability are unchanged. The Brooks 2009 entry is the \emph{input} shape from \demfile, computed by linear half-max interpolation; the entries marked ``recovered'' are \demreg{} pipeline outputs.}
\label{tab:shape}
\begin{tabular}{lrrrr}
\toprule
Source & Peak $\log T$ & $T_{\mathrm{peak}}$ (MK) & FWHM (in $\log T$) & Floor-corrected \\
\midrule
Isothermal Maxwellian (demregpy floor)   & 6.15  & 1.41 & 0.174          & ---   \\
$\kapp = 3$ recovered                    & 6.175 & 1.50 & 0.191          & 0.079 \\
\citet{brooks2009}, \demfile{} input     & 6.05  & 1.12 & 0.220          & ---   \\
$\kapp = 2.5$ single-T recovered         & 6.175 & 1.50 & 0.222 & 0.138 \\
$\kapp = 2.5$ multi-T (Brooks shape)     & 5.975 & 0.94 & 0.305          & 0.250 \\
Brooks-shape Maxwellian forward          & 5.95  & 0.89 & 0.319          & 0.267 \\
$\kapp = 2$ recovered                    & 6.025 & 1.06 & 0.353          & 0.307 \\
\bottomrule
\end{tabular}
\end{table}

\paragraph{FWHM derivation note.} FWHM $= 0.220$ was computed by linear half-max interpolation on the \citet{brooks2009} \demfile{} distributed file. The Brooks 2009 paper itself does not report FWHM; the paper represents the DEM via spline knots and characterizes its shape qualitatively. The Brooks input shape is the canonical EIS-derived QS DEM and is the published reference shape against which AIA-imaging analyses are commonly benchmarked, but it is not itself a regularized-inversion AIA-pipeline output; the inversion methods differ.

\paragraph{Matching-pipeline reference distribution.} A pipeline-matched comparison runs real quiet-Sun AIA observations through the same regularized inversion. Sampling 80 quiet patches across the disk on 2019 December 1 (solar minimum) at the standard AIA cadence, processing each through L1.5 calibration and the identical \demreg{} configuration used for the synthetic $\kapp$ test, the recovered FWHM distribution has median 0.283 (range 0.230--0.401) and median peak $\log T = 5.975$, stable across a second solar-minimum date (\S\ref{sec:comparison}). FWHM is defined by linear half-max interpolation on the regularized DEM curve in all entries of Table~\ref{tab:shape}.

Two multi-thermal synthetic sources sit inside this distribution. A multi-T $\kapp = 2.5$ source (the Brooks shape populated by Dz23 v10.1 $\kapp$ ion fractions across the temperature grid, with synthetic DN built from the per-ion $T_{\mathrm{eff}}$ checkpoint of \S\ref{sec:perion} factorized by the Maxwellian-rate-near-formation-T approximation) recovers FWHM $= 0.305$, peak $\log T = 5.98$, $\chisq/\dof = 0.17$ (the per-channel shot-noise model conservatively over-estimates uncertainty for a smoothly varying source matching the regularization's smoothness prior; sub-unity is the expected outcome, not a pathological fit), near the distribution median. The multi-T $\kapp$ source thus aligns \emph{both} shape metrics with the real-QS sample --- its recovered peak $\log T = 5.975$ equals the sample median and its FWHM sits inside the distribution --- where the single-T probe aligns width alone (its peak, $\log T = 6.175$, reflects the isothermal input rather than the sample). A Brooks-shape multi-T Maxwellian DEM forward-modeled through the same AIA channels returns FWHM $= 0.319$, also inside. The single-T $\kapp = 2.5$ source recovers FWHM $= 0.222$, just below the narrow edge: as a strictly isothermal pipeline probe, the fractional shortfall reflects the absence of multi-thermal broadening, not a failure of the convergence-principle prediction (\S\ref{sec:absorption}).

The $\kapp$ correction averages out under multi-T integration because each channel samples its dominant ion near formation temperature, where the $\kapp/$Maxwellian ion-fraction ratio is close to unity. The convergence-principle prediction holds across all three source families: single-T $\kapp$, multi-T Maxwellian, and multi-T $\kapp$ all produce recovered DEMs with FWHM in the 0.22--0.32 range, inside or just below the real-QS pipeline-output distribution (median 0.283, range 0.230--0.401), none of them distinguishable from real QS at the AIA-imaging level on FWHM alone. \S\ref{sec:break} returns to the structural degeneracy.

\paragraph{Reference-DEM robustness.} The degeneracy conclusion does not depend on the choice of the Brooks reference shape. Running the same forward-model-and-invert test on four additional published quiet-Sun DEMs (Table~\ref{tab:refdem}) --- the \citet{vernazza1978} average quiet Sun distributed with CHIANTI, the \citet{raymond1981} Skylab average quiet Sun (their Fig.~3, converted from binned emission measure to DEM per their Eq.~(1)), the \citet{warrenbrooks2009} off-limb Gaussian DEM (their Eq.~(3) with their printed best-fit parameters), and the near-isothermal off-disk quiet-Sun component of \citet{landi2008} (their Table~3, with their stated width $\Delta \log T \approx 0.05$) --- gives the same picture: for every broad reference shape the Maxwellian and $\kapp = 2.5$ recoveries differ by at most 0.018 dex --- the indistinguishability that carries the degeneracy claim. Three of the four references sit comfortably interior to the real-QS band; the broadest, \citet{raymond1981}, lands at the band's upper edge and is sensitive to the emission-measure-to-DEM conversion (Table~\ref{tab:refdem}), so it is read here as supporting the indistinguishability rather than band membership. The near-isothermal \citet{landi2008} reference is the informative exception: its Maxwellian forward model recovers the algorithmic floor (0.175, below anything the real-QS sample produces), while the same published shape under $\kapp = 2.5$ is absorbed as apparent multi-thermal structure inside the observed band (0.262) --- the single-T-probe behavior of Table~\ref{tab:shape} reproduced from a published reference curve.

\begin{table}[!htbp]
\centering
\caption{Reference-DEM robustness: recovered FWHM (in $\log T$) for a multi-thermal Maxwellian source and a multi-thermal $\kapp = 2.5$ source built on each published quiet-Sun DEM shape and run through the identical pipeline. The real-QS pipeline-output band is 0.230--0.401 (cross-date union 0.225--0.412). The Brooks row, recomputed within the shared robustness forward model, agrees with the Table~\ref{tab:shape} values (0.319, 0.305) to within 0.003 dex. For the \citet{raymond1981} shape, using the plotted axis values directly without the emission-measure-to-DEM conversion of their Eq.~(1) would give 0.522/0.409, bracketing the conversion sensitivity. The near-isothermal \citet{landi2008} row is the $\kapp$-signature case discussed in the text, not a degeneracy case.}
\label{tab:refdem}

\begin{tabular}{llrrr}
\toprule
Reference DEM & Type & FWHM (Mxw) & FWHM ($\kapp = 2.5$) & $\Delta$ \\
\midrule
\citet{brooks2009} \demfile          & on-disk EIS QS            & 0.322 & 0.307 & $-0.015$ \\
\citet{vernazza1978} (CHIANTI)       & on-disk average QS        & 0.387 & 0.378 & $-0.009$ \\
\citet{raymond1981} Fig.~3           & on-disk average QS        & 0.412 & 0.399 & $-0.013$ \\
\citet{warrenbrooks2009} Gaussian    & off-limb EIS QS           & 0.338 & 0.320 & $-0.018$ \\
\citet{landi2008} QS component       & off-disk, near-isothermal & 0.175 & 0.262 & $+0.087$ \\
\bottomrule
\end{tabular}

\end{table}

\paragraph{Influence of the AIA pipeline floor.} \citet{guennou2012a,guennou2012b} document that AIA-pipeline DEM solutions cluster near a width set by the AIA response-function structure and the smoothing applied by the specific inversion method (parametric Gaussian-fit in their case, regularized Tikhonov here) rather than by underlying physics. We measure the floor directly for the present pipeline: a synthetic isothermal Maxwellian DEM at $\log T = 6.176$ (no $\kapp$ modification) run through the same \demreg{} pipeline recovers FWHM $= 0.174$ at peak $\log T = 6.15$ --- the irreducible width contributed by the AIA response kernels plus GSVD-selected regularization, against which all physical broadening is measured (Table~\ref{tab:shape}). A floor of pipeline-mechanical origin would not vary with $\kapp$ because the smoothing scale set by GSVD is independent of the physics of the synthetic input; our recovered FWHM does vary monotonically (0.191, 0.222, 0.353 at $\kapp = 3, 2.5, 2$, all above the 0.174 floor). The $\kapp$-driven broadening above the floor at $\kapp = 2.5$ is $\sim 0.05$ dex (simple differences of raw widths; the quadrature-corrected equivalents are the floor-corrected column of Table~\ref{tab:shape}, 0.138 for the single-T probe); the multi-T integration adds $\sim 0.13$ dex; in combination the recovered FWHM sits inside the real-QS pipeline-output distribution but does not by itself uniquely identify the underlying source.

The peak offset ($\Delta\log T = 0.125$ between $\kapp = 2.5$ and Brooks's reference shape) reflects the choice of single-input $\Teff$ against the real quiet Sun's integration over a distribution of structures, and the recovered peak from the real-QS demregpy distribution clusters at $\log T \approx 5.97$ rather than 6.18. Peak location depends on input parameters; FWHM is the meaningful shape metric, and the matching-pipeline test shows the FWHM is itself a degenerate metric across the source families demregpy admits.

All recovered DEMs show a bimodal structure~(Fig.~\ref{fig:sensitivity}): a main coronal peak at $\log T \approx 6.0$--6.2, a gap near 6.4, and a secondary component at 6.7--6.8 from Fe XVI--XVII tail emission. \S\ref{sec:hottail} returns to this point.

\begin{figure}[t]
\centering
\includegraphics[width=0.85\linewidth]{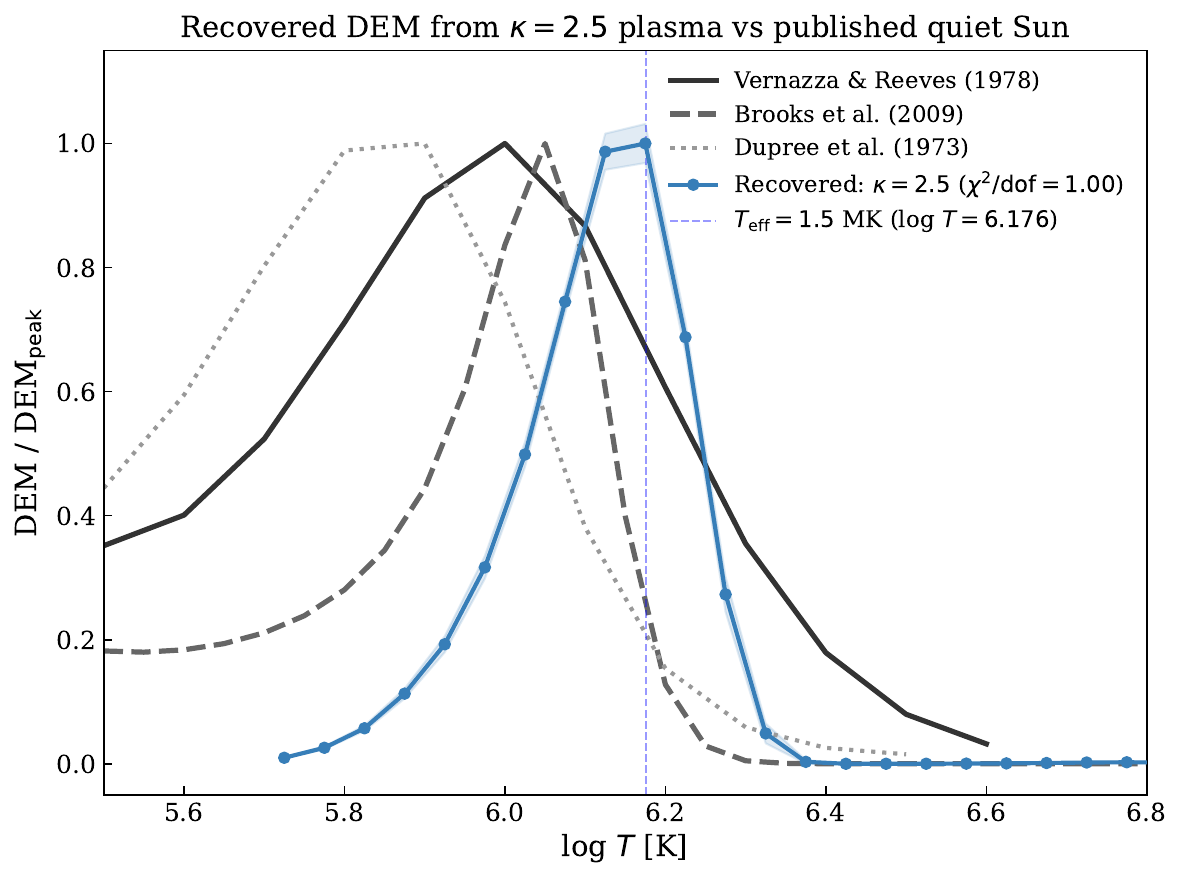}
\caption{Headline figure. Normalized DEM overlay: $\kapp = 2.5$ recovered (with regularization error envelope) vs.\ \citet{brooks2009} \demfile{} reference. The FWHM of the two curves: 0.222 (single-T $\kapp$ recovered) and 0.220 (Brooks-derived input shape). Two further published quiet-Sun DEMs are overlaid for context: the CHIANTI v11 default quiet-Sun DEM, derived from the \citet{vernazza1978} average quiet Sun, and the CHIANTI v3 quiet-Sun DEM derived from OSO-6 spectra \citep{dupree1973}. Both are full-atmosphere DEMs whose maxima are chromospheric; each published curve is therefore normalized to its coronal peak within the plotted window. The controlled reference-DEM comparison is Table~\ref{tab:refdem} (\S\ref{sec:demshape}). The single-T recovered FWHM at $\chisq/\dof = 1.00$ sits just below the narrow edge of the real-QS demregpy pipeline-output distribution (median 0.283, range 0.230--0.401 across 80 quiet patches in AIA observations, stable across two minimum dates) as a strictly isothermal pipeline probe; a multi-T $\kapp$ source built from the same Brooks shape recovers FWHM $= 0.305$, inside the distribution near the median (Table~\ref{tab:shape}, \S\ref{sec:demshape}). The absolute FWHM does not by itself distinguish a $\kapp$ source from a multi-T Maxwellian source through this pipeline.}
\label{fig:headline}
\end{figure}

\begin{figure}[t]
\centering
\includegraphics[width=0.95\linewidth]{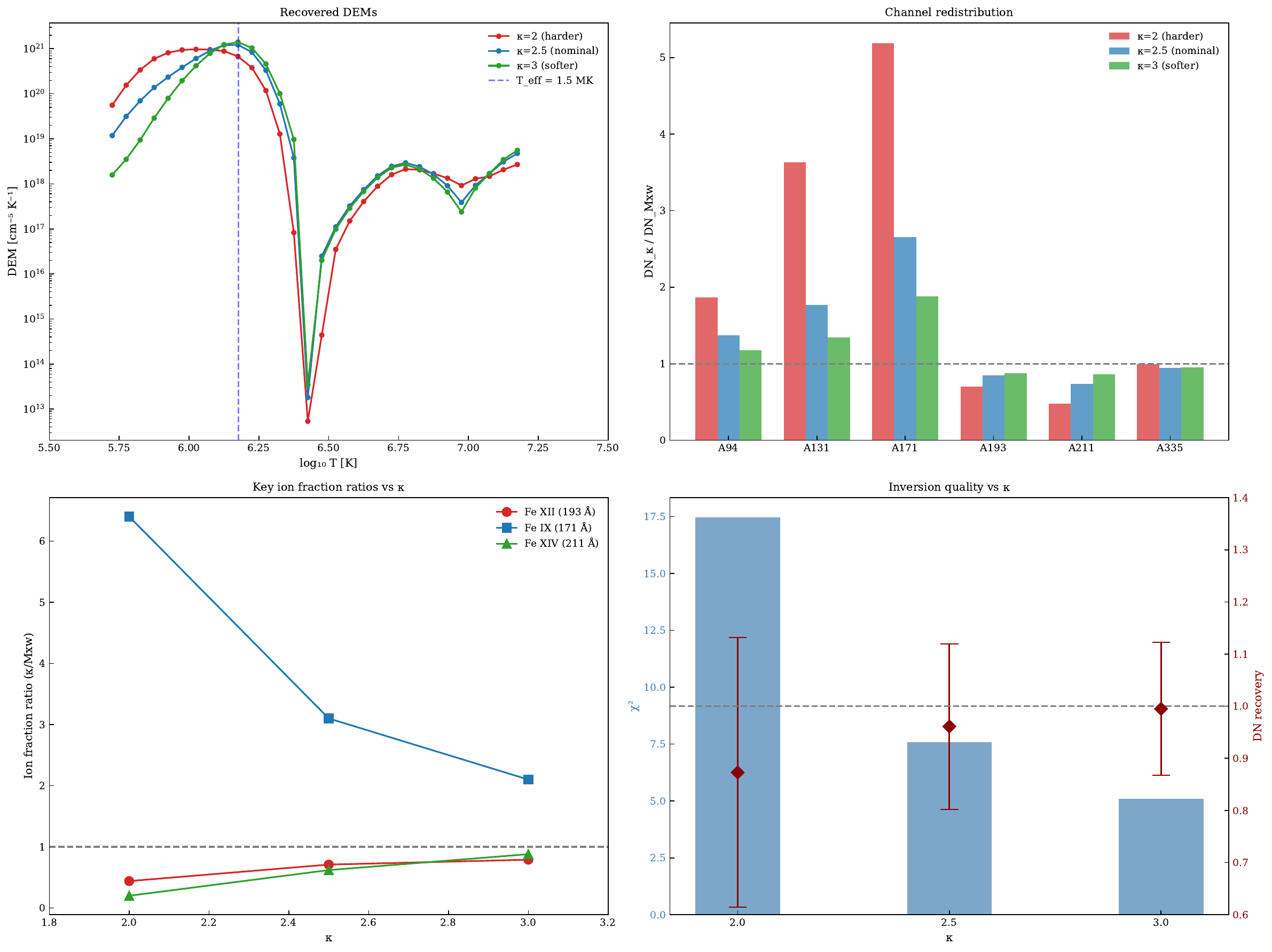}
\caption{$\kapp$ sensitivity comparison. Four panels: (a) recovered DEMs for $\kapp = 2, 2.5, 3$; (b) per-channel $\kapp$/Mxw DN ratios; (c) ion-fraction ratios ($\kapp$/Mxw) versus $\kapp$ for the key diagnostic ions Fe IX, Fe XII, Fe XIV; (d) inversion quality versus $\kapp$: raw lines-only $\chisq$ (bars, left axis) with the per-channel DN-recovery range (markers, right axis).}
\label{fig:sensitivity}
\end{figure}

%------------------------------------------------------------------
\subsection{Abundance and continuum sensitivity}\label{sec:sensitivity}
%------------------------------------------------------------------

Coronal vs.\ photospheric abundances at $\kapp = 2.5$: $\Delta\chisq/\dof < 0.01$. The only measurable channel-level difference is in 131 \AA{} (ratio 1.771 coronal vs.\ 1.527 photospheric) from FIP-sensitive non-Fe contributions. The DEM result is insensitive to the assumed abundance model.

Including continuum improves the fit from $\chisq/\dof = 1.52$ to $1.00$ at $\kapp = 2.5$. The lines-only analysis is the conservative case. Continuum primarily affects 131 \AA{} (23\% fraction) and is negligible ($<1\%$) elsewhere. Under the per-ion free-bound treatment of \S\ref{sec:continuum} the continuum-included fit is confirmed directly: $\chisq/\dof = 0.99$--1.00 across the computed range of treatments, with peak $\log T$ and FWHM unchanged.

%------------------------------------------------------------------
\subsection{Emission-measure self-consistency with Brooks 2009 absolute radiances}\label{sec:emrec}
%------------------------------------------------------------------

\S\ref{sec:demshape} shows the recovered DEM \emph{shape} sits inside the real-QS pipeline-output distribution. Two questions follow at the absolute-radiance level. First, is the framework self-consistent: does the emission measure required for the $\kapp$ source to reproduce the actual line radiances \citet{brooks2009} measured sit inside the standard quiet-Sun white-light EM range? If not, the shape match is a coincidence rather than a physical signature. Second, does a single-T $\kapp$ source forced to reproduce real multi-T radiances display a specific spectroscopic signature? It does: a monotonic EM tilt with ion stage, because hot ions and cool ions sample different parts of the actual multi-T plasma but must be reconciled at a single source temperature.

\paragraph{Physical origin of the predicted EM tilt.} A single-T source at $\Teff = 1.5$ MK reproduces line radiances by adjusting the assumed EM. Under the Maxwellian Brooks DEM, hot ions (Fe XIV, XV, XVI) form preferentially in regions where the actual plasma sits at higher temperatures, contributing radiances that an isothermal source at $\Teff = 1.5$ MK can match only by reducing its assumed EM. Cool ions (Fe IX, X) form preferentially in regions where the actual plasma sits at lower temperatures, contributing radiances that the isothermal source can match only by increasing its assumed EM. The $\kapp$ correction reweights ion fractions but does not fix the underlying single-T limitation, so the EM required to match cool-ion radiance must be higher than the EM required to match hot-ion radiance. The tilt is monotonic in ion stage; its magnitude reflects how multi-thermal the actual line of sight is.

For each of eight strong Fe IX--XVI EUV coronal lines spanning the formation-temperature range of the \citet{brooks2009} DEM (most observed by Hinode/EIS; Fe IX 171.07 \AA{} and Fe XVI 335.41 \AA{} canonically observed by SDO/AIA and SOHO/CDS rather than EIS), we compute: (i) the line radiance predicted by the Brooks DEM under Maxwellian physics; (ii) the emissivity at $\Teff = 1.5$ MK under $\kapp = 2.5$, constructed from the Maxwellian emissivity times $f_{\mathrm{ion}}^{\kapp}/f_{\mathrm{ion}}^{\mathrm{Mxw}}$; and (iii) the EM required for a single-T $\kapp$ source at $\Teff = 1.5$ MK to reproduce the Brooks-predicted radiance. Density $n_{e} = 10^{9}$ cm$^{-3}$; coronal abundances; CHIANTI v11.0.2.

\begin{table}[!htbp]
\centering
\caption{Line-by-line EM required for $\kapp = 2.5$ at $\Teff = 1.5$ MK to reproduce the \citet{brooks2009} predicted radiances of eight strong Fe IX--XVI EUV coronal lines. Fe IX 171.07 \AA{} sits at the edge of the Hinode/EIS short-wavelength band, also observed by SDO/AIA and SOHO/CDS; Fe XVI 335.41 \AA{} is outside the EIS bandpass entirely and is observed by SDO/AIA (335 channel) and SOHO/CDS. Values are monotonically decreasing across the ion-stage sequence, the predicted signature of $\kapp$ ion-fraction reweighting against a single-T source temperature.}
\label{tab:emrec}
\begin{tabular}{lrrrrr}
\toprule
Line & $\lambda$ (\AA) & $f_{\kapp}/f_{\mathrm{Mxw}}$ & $R_{\mathrm{Brooks}}$ (erg cm$^{-2}$ s$^{-1}$ sr$^{-1}$) & EM$_{\kapp}$ (cm$^{-5}$) & $\log\,$EM$_{\kapp}$ \\
\midrule
Fe IX   & 171.07 & 3.10 & $5.79\times 10^{2}$  & $8.48\times 10^{26}$ & 26.93 \\
Fe X    & 184.54 & 1.73 & $6.45\times 10^{1}$  & $3.78\times 10^{26}$ & 26.58 \\
Fe XI   & 188.22 & 1.04 & $1.23\times 10^{2}$  & $2.22\times 10^{26}$ & 26.35 \\
Fe XII  & 195.12 & 0.71 & $1.03\times 10^{2}$  & $1.55\times 10^{26}$ & 26.19 \\
Fe XIII & 202.04 & 0.63 & $2.70\times 10^{1}$  & $1.05\times 10^{26}$ & 26.02 \\
Fe XIV  & 264.79 & 0.62 & $6.06$               & $8.73\times 10^{25}$ & 25.94 \\
Fe XV   & 284.16 & 0.71 & $9.49$               & $8.56\times 10^{25}$ & 25.93 \\
Fe XVI  & 335.41 & 1.40 & $9.62\times 10^{-1}$ & $7.45\times 10^{25}$ & 25.87 \\
\bottomrule
\end{tabular}
\end{table}

\paragraph{Result.} Median $\log\,\mathrm{EM}_{\kapp} = 26.10$; range $[25.87, 26.93]$. All values sit within the standard quiet-Sun white-light EM range ($10^{26}$--$10^{28}$ cm$^{-5}$). The Brooks integrated DEM is $\log \int \mathrm{DEM}\, dT = 26.50$, in the same range. The framework is self-consistent with white-light electron-density observations.

\paragraph{The systematic spread is the single-T idealization signature.} The 1.06 dex spread is not random. Hot ions (Fe XIV, XV, XVI) require $\log\,\mathrm{EM} \approx 25.9$; cool ions (Fe IX, X) require $\log\,\mathrm{EM} \approx 26.6$--26.9. Tight EM-loci convergence is the characteristic signature of genuinely isothermal coronal regions; off-limb QS analyses report DEMs ``strongly peaked'' near 1 MK with narrow distributions \citep{warrenbrooks2009}, while multi-thermal QS observations with line-of-sight integration show wider scatter consistent with the intrinsic multi-T structure \citep{lorincik2020}. The 1.06 dex tilt here is larger than the typical multi-T scatter because the test deliberately fits a strictly single-T idealization against multi-T radiances, and the residual stacks the multi-T spread on top of the $\kapp$ ion-fraction reweighting. Because the comparison is restricted to lines of a single low-FIP element (Fe IX--XVI), the tilt is immune to FIP-bias confounding: the coronal abundance factor is common to every line and cancels in the EM-loci ratio, so spatial variation in the FIP enhancement cannot masquerade as a temperature tilt. A realistic multi-thermal $\kapp$ plasma along the line of sight would resolve at the EIS spectroscopic level with per-ion EM loci converging at reduced spread.

The signed direction of the tilt is the $\kapp$-specific structural content, separable in principle from a multi-T Maxwellian: ion-fraction reweighting under $\kapp$ enhances cool ions and suppresses hot ions, so the EM required to reproduce cool-ion radiances at fixed source temperature is lower and hot-ion EM is higher. EIS spectroscopy actively resolves what AIA imaging cannot: within-ion line ratios sample temperature inside an ion's formation range and break the AIA-imaging degeneracy at the channel-integration level \citep{dudik2014b,lorincik2020}. Disentangling $\kapp$-tilt from multi-T scatter observationally requires forward-modeling $\kapp$-modified ion fractions on top of a prescribed multi-T DEM and comparing against EIS spectroscopy; \S\ref{sec:openleft} flags the relevant compute.

The multi-T $\kapp$ source already constructed in \S\ref{sec:demshape} predicts where the EM-loci spread should land. Each ion in that source samples its own formation temperature, where the $\kapp$/Maxwellian ion-fraction ratio is close to unity, so the per-ion EM-loci spread reduces to the multi-T scatter of the underlying DEM \citep[the wider scatter typical of multi-T QS analyses;][]{lorincik2020} plus a residual signed component of order $(\kapp - 3/2)/\kapp$ riding on top of it. Direct compute confirms the collapse: weighting each ion's fraction by the Brooks DEM and propagating to the per-line EM-loci ratio gives a residual spread of 0.14 dex across the same eight Fe IX--XVI lines, 14\% of the 1.06 dex single-T amplitude and below the $\sim 0.2$ dex multi-T scatter typical of QS EIS analyses \citep{lorincik2020}. A sensitivity check shows that Fe XVI 335.41 \AA{}, which forms at $\log T \approx 6.4$ where the Brooks DEM has minimal weight, contributes disproportionately to the residual: dropping Fe XVI from the line set reduces the spread to 0.10 dex (9.5\% of the single-T amplitude). The collapse range 0.10--0.14 dex is robustly below the multi-T QS scatter regardless of whether Fe XVI is included. The 1.06 dex single-T tilt of Table~\ref{tab:emrec} sets an upper bound on the $\kapp$ residual one would extract from a multi-T EIS analysis, not a prediction of the spread itself.

\begin{figure}[t]
\centering
\includegraphics[width=0.8\linewidth]{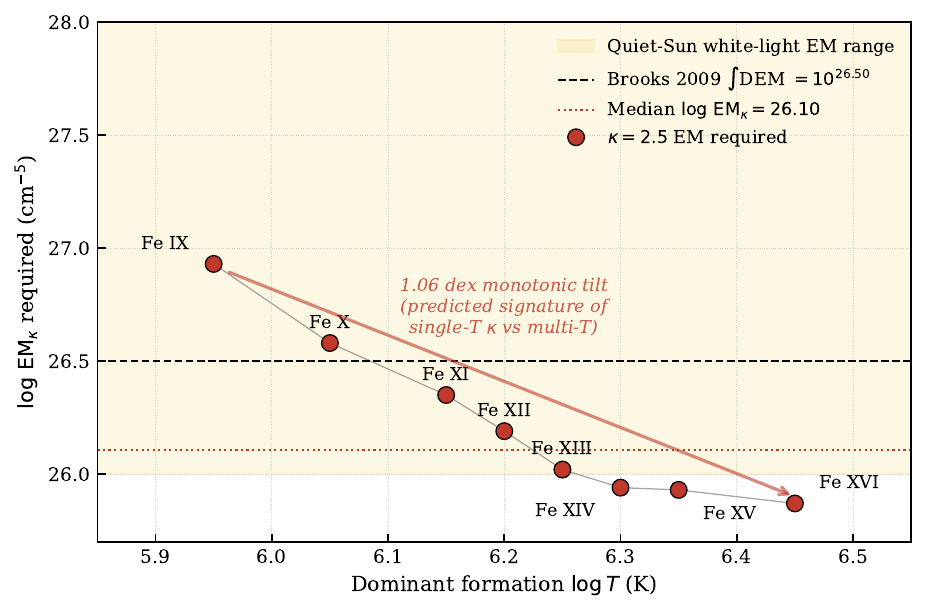}
\caption{EM reconciliation across ion stages. $\log\,\mathrm{EM}_{\kapp}$ vs.\ dominant formation $\log T$ for the eight Fe IX--XVI EUV coronal lines of Table~\ref{tab:emrec}. The Brooks integrated DEM ($\log = 26.50$) and the standard quiet-Sun white-light EM range ($10^{26}$--$10^{28}$ cm$^{-5}$) are marked. The monotonic tilt across ion stages is the predicted signature of a single-T $\kapp$ source confronting real multi-T radiances.}
\label{fig:emrec}
\end{figure}

The $\kapp$ signature exists at the per-line EM-loci level~(Fig.~\ref{fig:emrec}) but is observationally confounded by the wider multi-thermal scatter typical of real line-of-sight integration. Its empirical detection requires extracting a signed signal from that multi-thermal scatter rather than reading a standalone tilt.

The inverse conclusion is also useful to state directly. Published quiet-Sun EUV-spectroscopy analyses of these ions show spreads consistent with the underlying multi-thermal DEM width and do not routinely exhibit a clean 1.06 dex monotonic tilt across Fe IX--XVI. A strictly isothermal $\kapp$ plasma is ruled out as a model of the real quiet Sun: the observed spread is too narrow for the single-T idealization. The real quiet corona must possess genuine multi-thermal structure along the line of sight; the present paper's claim is that the AIA-imaging pipeline cannot \emph{distinguish} multi-T Maxwellian from multi-T $\kapp$ from intermediate combinations, not that the corona is mono-$\kapp$.

%------------------------------------------------------------------
\subsection{The EUV continuum reversal}\label{sec:euvreversal}
%------------------------------------------------------------------

The common intuition: $\kapp$ distributions produce uniformly ``harder'' continuum spectra. This is correct for X-rays ($E \gg k\Teff$), where the power-law tail dominates. At EUV wavelengths ($E \leq k\Teff$), the opposite occurs: the $\kapp$ distribution depletes electrons near the thermal energy, redistributing them into the far suprathermal tail, and at $E \approx k\Teff$ this depletion reduces free-free emission below the Maxwellian level.

Applying Equation~(\ref{eq:ffratio}) at AIA central wavelengths with $\kapp = 2.5$, $\Teff = 1.5$ MK: the ratio is 0.785 at 94 \AA{} ($E = 132$ eV $\approx k\Teff$, a 22\% suppression), 1.005 at 131 \AA, 1.235 at 171 \AA, 1.329 at 193 \AA, 1.405 at 211 \AA, and 1.802 at 335 \AA{} ($E = 37$ eV, an 80\% enhancement). The crossover lies near 131 \AA{} where $E \approx k\Teff$. The original $\kapp$ free-free derivation appears in \citet{dudik2012}; the AIA-channel-level quantification is reported here for the first time. Any analysis that models $\kapp$ continuum in the EUV should account for this energy-dependent behavior.

%==================================================================
\FloatBarrier

\section{Discussion}\label{sec:discussion}
%==================================================================

The discussion has two parts. \S\ref{sec:absorption}--\S\ref{sec:demwidth} interpret the diagnostic degeneracy itself: the multi-thermal-vs-mono-$\kapp$ ambiguity, the hot DEM tail, and DEM width as a $\kapp$ diagnostic. \S\ref{sec:closure} applies the moment-mismatch implication to the standard quiet-Sun energy budget.

%------------------------------------------------------------------
\subsection{Absorption of the kappa signature by regularized inversion}\label{sec:absorption}
%------------------------------------------------------------------

The DEM inversion does not reject $\kapp$ emission; it accommodates it. The regularized inversion has enough degrees of freedom (31 temperature bins for 6 channel constraints) to redistribute the DEM shape until the Maxwellian-predicted DN match the $\kapp$-generated DN. This is a consequence of the problem being under-determined, not a failure of the algorithm. The inversion finds the smoothest non-negative DEM consistent with the data, and the $\kapp$ channel redistribution is smooth enough to be absorbed.

The result is a fundamental degeneracy at the AIA-imaging pipeline-output level: a multi-thermal Maxwellian DEM and a single-$\kapp$ plasma produce the same AIA DN values to within $\chisq/\dof = 1.00$. In the language of \S\ref{sec:intro}~\textparagraph 4, the coarse-graining is closed---the variable being inferred does not resolve the variable being interpreted.

The degeneracy is specific to the AIA-imaging level. Full Hinode/EIS spectroscopy with within-ion line ratios can break it: \citet{dudik2014b} for Fe IX--XIII and \citet{dudik2019} for Fe XVII--XVIII identify specific within-ion line ratios that distinguish $\kapp$ from multi-T at the spectroscopic level, and \citet{lorincik2020} finds quiet-Sun EIS spectra consistent with Maxwellian. The two findings are consistent with the convergence principle: any ionization-gated diagnostic returns $\Teff$, but spectroscopy and imaging differ in how they compress the within-ion information that survives the ionization gate. AIA channel integration collapses information that EIS preserves.

%------------------------------------------------------------------
\subsection{Implications for the DEM hot tail}\label{sec:hottail}
%------------------------------------------------------------------

The secondary hot component at $\log T \approx 6.7$--6.8 in the recovered DEM arises from Fe XVI--XVII enhancement under $\kapp$ (ratios 1.4$\times$ and 6.5$\times$ at $\kapp = 2.5$, Table~\ref{tab:feratios}). The inversion attributes this enhanced charge-state population to a separate hot plasma component. This is the same ``hot tail'' that appears in published quiet-Sun DEMs from AIA-imaging analysis and is commonly interpreted as evidence of nanoflare heating.

The hot tail does not rule in or rule out nanoflare heating in the quiet Sun. It introduces an ambiguity at the AIA-imaging level: the hot DEM tail in QS data is consistent with genuine hot plasma from impulsive heating, with $\kapp$ tail emission from a single distribution, or with a combination. DEM-slope methodology, well-developed for active regions \citep{warren2012,cargill2014} and applied analogously to QS data, anchors constraints on impulsive-heating frequency distributions that should be evaluated in light of this ambiguity when the underlying source distribution is non-Maxwellian.

\paragraph{Scope note.} Active-region direct hot-plasma detections---FOXSI-2 $\geq$10 MK in non-flaring AR \citep{ishikawa2017}; NuSTAR upper limits in non-flaring AR \citep{hannah2016}; pervasive Fe XIX 592.2 \AA{} in AR \citep{brosius2014}---are not contradicted by this section. They sample the AR regime, where impulsive driving and high density dominate the relevant physics. The \S\ref{sec:hottail} ambiguity claim is about the QS hot DEM tail specifically. Per the regime distinction in \citet{edmonds2026a}, AR loops are dynamic, AR cores are quasi-steady, QS is its own regime. The \S\ref{sec:nanoflarepillars} nanoflare-pillar discussion respects the same scope.

%------------------------------------------------------------------
\subsection{Degeneracy and what breaks it}\label{sec:break}
%------------------------------------------------------------------

The multi-thermal / mono-$\kapp$ degeneracy at the AIA-imaging level cannot be resolved by adding more channels at the same diagnostic level. All AIA-imaging channels are downstream of the same ionization gate, and the \citet{hannah2012}/\demreg{} framework collapses them at the same coarseness. The degeneracy is likewise independent of the inversion algorithm: the within-ion information is removed at channel integration, before any inversion runs, so no algorithm consuming the six AIA integrals under a Maxwellian response --- regularized, MCMC, or sparse --- can recover what the integration has already discarded. The regularization sweep of \S\ref{sec:dem} is a within-framework confirmation of this structural fact. The within-ion line-ratio information that does break the degeneracy \citep{dudik2014b,dudik2019} is preserved at the spectroscopic level (EIS, IRIS) and destroyed at the imaging level (AIA channel integrals).

This is the AIA-imaging-specific instance of the convergence principle: ionization-based diagnostics converge on $\Teff$ because they sit downstream of a single threshold-crossing bottleneck in the ionization-recombination balance. The quiet-corona EUV-detection convergence addressed in \citet{edmonds2026a} \S 4.6 (five independent EUV methods returning $\Teff \approx 1.5$ MK) is explained by the same bottleneck; the AIA-imaging DEM-inversion pipeline is one more method that convergence absorbs.

The degeneracy can in principle be broken by diagnostics that sample the distribution core rather than the tail: radio bremsstrahlung (where $T_{\mathrm{brightness}} \propto \Tcore$ at optically thin frequencies), Thomson scattering (sensitive to the bulk electron population), or direct spectral measurements at energies where the $\kapp$ power law diverges from exponential decay (soft X-rays). Within EUV, full-line-resolution spectroscopy (EIS, IRIS) preserves enough within-ion structure to constrain $\kapp$; AIA-imaging analysis does not.

\paragraph{A concrete EIS validation test.} The sharpest archival test uses the Fe IX ratio--ratio diagnostic identified by \citet{dudik2014b} (their \S 6.1 and Fig.~8): the plane of 177.592/171.073 \AA{} against 189.941/177.592 \AA. The diagnostic is within-ion --- its $\kapp$ sensitivity originates in excitation alone, so it is immune to the ionization-equilibrium gating that drives the convergence-principle degeneracy --- has the lowest density sensitivity of the Fe IX combinations, and separates $\kapp = 2$ from Maxwellian by approximately a factor of 2, well above EIS photometric uncertainty for these strong lines. All three lines fall in the EIS short-wavelength band, so archival quiet-Sun rasters suffice; \citet{dudik2014b} flag Fe IX 197.862 \AA{} and the Fe X 365.560/180.441 \AA{} pair as alternates. The expected departure at $\kapp = 2.5$ is below the factor-2 value at $\kapp = 2$ but of the same order. The existing within-ion quiet-Sun result of \citet{lorincik2020} is Maxwellian-consistent; how strongly that null constrains $\kapp = 2.5$ depends on the $\kapp$-sensitivity of the specific diagnostics employed there, which were not selected for it, and quantifying the $\sigma$-weight of the existing null against the proposed diagnostic is part of the proposed analysis itself. A dedicated ratio--ratio analysis at $\kapp = 2.5$ sensitivity on deep quiet-Sun rasters is the discriminating measurement that the AIA-level result of this paper cannot supply. The diagnostic also survives line-of-sight multi-thermal integration in a way the EM-loci tilt of \S\ref{sec:emrec} does not: all three lines belong to a single ion, so they share one formation window and one ionization gate, and the residual systematic is the temperature variation within the Fe IX window --- precisely what the temperature--$\kapp$ ratio--ratio diagrams of \citet{dudik2014b} parameterize. What deep-raster archives may still lack is photon throughput: spatiotemporal averaging over many quiet-Sun network elements can dilute a real $\kapp$ signature, and if archival EIS statistics prove marginal at the $\kapp = 2.5$ level, the higher-throughput next-generation EUV spectrographs (MUSE, Solar-C/EUVST) are the instruments that would settle it.

This does not establish that the quiet Sun is mono-$\kapp$ --- real coronal structure is spatially complex --- only that AIA-imaging DEM analysis alone cannot uniquely constrain thermal structure when non-Maxwellian distributions are admissible.

%------------------------------------------------------------------
\subsection{DEM width as a kappa constraint}\label{sec:demwidth}
%------------------------------------------------------------------

Lower $\kapp$ produces a broader recovered DEM: FWHM = 0.191 ($\kapp = 3$), 0.222 ($\kapp = 2.5$), 0.353 ($\kapp = 2$). The relationship is monotonic and follows directly from the charge-state broadening of \S\ref{sec:fexi}. The Brooks-derived FWHM = 0.220 (computed from \demfile) is consistent with $\kapp \approx 2.5$ and inconsistent with $\kapp = 2$ (too broad) or $\kapp \to \infty$ (zero width from isothermal Maxwellian).

If the AIA-imaging recovered FWHM for the quiet Sun is partly or wholly attributable to $\kapp$ rather than thermal structure, it admits an alternative reading: rather than constraining the range of temperatures present in the corona, the DEM width constrains the degree of departure from Maxwellian.

%------------------------------------------------------------------
\subsection{The closure problem behind the standard quiet-Sun energy budget}\label{sec:closure}
%------------------------------------------------------------------

\subsubsection{The closure problem}\label{sec:closure-problem}

The standard quiet-Sun energy-budget calculation asks how much heat the corona loses and how much energy must be deposited to maintain the temperatures we observe. The dominant loss term is conductive: heat conducted along magnetic field lines from the hot corona down toward the transition region and chromosphere. Conduction needs a temperature gradient and a conductivity. The conductivity is Spitzer-H\"arm:
\begin{equation}
q_{\mathrm{SH}} = -\kappa_{0}\, T^{5/2}\, \nabla T.
\label{eq:spitzer}
\end{equation}

\paragraph{Cross-discipline note on Spitzer-H\"arm.} Equation~(\ref{eq:spitzer}) is the leading-order Chapman-Enskog closure of the electron heat-flux moment hierarchy \citep{spitzer1953,braginskii1965,helander2002}. Like every moment closure, it assumes the electron distribution belongs to a specific family---Maxwellian---and that the temperature parameter of that family is the moment governing bulk transport. The formula's output is precise for that physics; the only freedom in its application is the temperature input. Standard quiet-Sun budget calculations use $T \approx 1.5$ MK, the temperature returned by EUV diagnostics.

\paragraph{Closure hierarchy and the extended Knudsen layer.} In the moment-hierarchy language of \citet{grad1949}, Spitzer-H\"arm is the leading-order truncation of the 13-moment expansion at vanishing Knudsen number; alternative closures (Grad-13 itself, Knudsen-layer matching, Levermore moment hierarchies) extend the expansion to finite $\mathrm{Kn}$ but assume the same near-Maxwellian distribution family at every order \citep{struchtrup2005}. The quiet corona at $\mathrm{Kn} \approx 0.01$--0.1 is in this sense an extended Knudsen layer: the local fluid closure fails not only at the chromospheric interface but across the bulk, because the underlying distribution sits outside the family any local moment closure can accommodate.

Under the $\kapp$ hypothesis (consistent with the AIA-DEM pipeline test of \S\ref{sec:results}), neither closure assumption holds. The electron distribution does not lie in the Maxwellian family. $\Teff$ is the moment that ionization-gated EUV diagnostics return for a $\kapp \approx 2.5$ distribution; it is dominated by the suprathermal tail, and the convergence principle extends this to any ionization-gated diagnostic at Knudsen number $> 0.01$. The bulk Maxwellian-like core of the $\kapp$ distribution sits at $\Tcore = (\kapp - 3/2)/\kapp \cdot \Teff \approx 0.6$ MK. Bulk transport is governed by the bulk. Spitzer-H\"arm with $\Teff$ as input is a Maxwellian-closure formula evaluated at a tail-weighted moment of a non-Maxwellian distribution.

\subsubsection{Failure of the local conductive closure}\label{sec:closure-nonexistence}

\begin{figure}[H]
\centering
\includegraphics[width=0.95\linewidth]{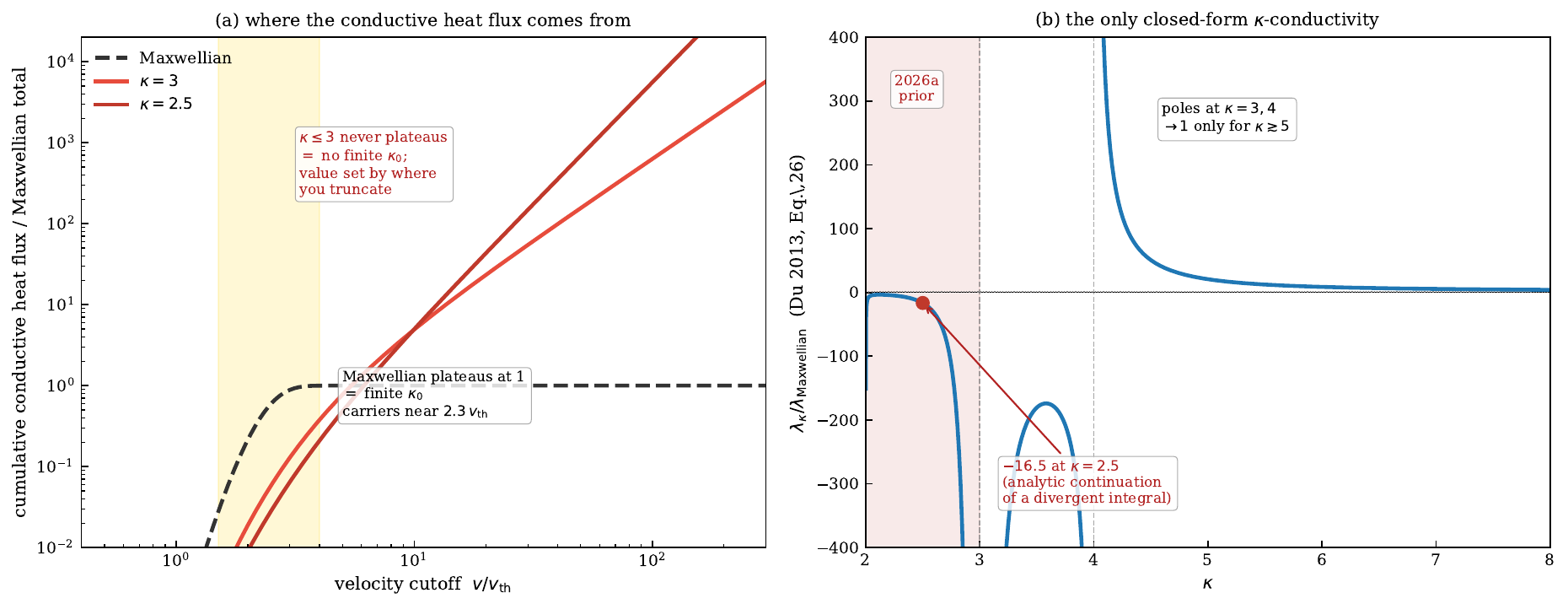}
\caption{The collapse of the local Spitzer-H\"arm closure under $\kapp$. \textbf{(a)} Cumulative conductive heat flux as a function of the upper velocity cutoff, normalized to the Maxwellian total, for the dominant heat-carrying term of the Lorentz conductivity \citep[Du][Eq.~25]{du2013}. The Maxwellian curve plateaus---a finite $\kappa_{0}$, with carriers near $2.3\,v_{\mathrm{th}}$---while the $\kapp = 2.5$ and $\kapp = 3$ curves never plateau: the integral does not converge, so no finite local conductivity exists and any value extracted is set by where the integral is truncated. \textbf{(b)} The only closed-form $\kapp$-conductivity \citep[Du][Eq.~26]{du2013}, ratio to the Maxwellian value: poles at $\kapp = 3$ and $\kapp = 4$, relaxing to unity only for $\kapp \gtrsim 5$. Across the 2026a range $\kapp \in [2,3]$ (shaded) it does not yield a finite positive conductivity; the finite $-16.5$ it returns at $\kapp = 2.5$ is an analytic continuation of the divergent integral of panel (a), not a physical value. Convention note: the figure uses the Lorentz ($e$--$i$) form for transparency; the $e$--$e$-inclusive \citep{guodu2019} and standard-$\kapp$ Boltzmann \citep{husidic2021} treatments carry the same poles, and the regularized-$\kapp$ form \citep{husidic2022} is finite but cutoff-dependent and $\sim 10^{4}\times$ Maxwellian. The locked values and the machine-precision validation gate are in the public analysis repository.}
\label{fig:closure}
\end{figure}

Conduction is the third velocity moment, and the Chapman--Enskog conductivity integrand carries the $v^{4}$ mean-free-path weighting of Coulomb collisions ($\nu \propto v^{-3}$); the conductive flux is therefore carried not by the bulk but by suprathermal electrons several thermal speeds out (Fig.~\ref{fig:closure}a). For a Maxwellian this is harmless: those carriers belong to a population a single temperature describes, the integral converges, and its carriers peak near $2.3\,v_{\mathrm{th}}$---which is exactly why Spitzer-H\"arm is valid for a Maxwellian. For the $\kapp$ distribution of the adopted premise it is fatal. A hierarchy of moment divergences runs along the $\kapp$ axis \citep{lazarfichtner2021}: at $\kapp \leq 3/2$ the temperature integral diverges and $\Teff$ itself is undefined; at $\kapp \leq 2$ the third velocity moment, the heat flux, diverges; at $\kapp \leq 5/2$ the fourth velocity moment, which governs the next-order Chapman-Enskog correction, diverges.

For the standard $\kapp$ distribution the local Spitzer-H\"arm conductivity diverges across the entire $\kapp$ range: the closed forms carry poles at $\kapp = 3$ and $\kapp = 4$ \citep{du2013,husidic2021}, the $e$--$e$-inclusive treatment of \citet{guodu2019} remains pathological at small $\kapp$ (its coefficient turns negative for $\kapp < 10$), and the only finite expression---for the regularized $\kapp$ distribution---is cutoff-dependent (scaling as $\xi^{-7}$ in the cutoff $\xi$) and of order $10^{4}$ times the Maxwellian value \citep{husidic2022}. In neither case is there a finite, convention-independent, order-unity local conductivity $\kappa_{0}$ for a temperature to be substituted into. The physical heat flux that does exist for $\kapp > 2$ is non-local: it depends on the coronal temperature over the long mean-free-path range of the tail carriers, not on the local gradient, so it is the Fourier-law form $q = -\kappa_{0}\,\nabla T$ itself---not the finiteness of the flux---that fails. The pathology is structurally the Burnett / super-Burnett asymptotic divergence of hard-sphere kinetic theory \citep{grad1949,struchtrup2005}: the Chapman-Enskog expansion is asymptotic rather than convergent and cannot be continued order-by-order on the structural boundary the $\kapp \approx 2.5$ regime occupies. There is no local fluid closure to correct at $\kapp = 2.5$; there is only the question of what the true, fully kinetic flux is---a question outside the scope of any moment closure, taken up in \S\ref{sec:opentransport}.

\subsubsection{The temperature-substitution trap}\label{sec:closure-correction}

The result of \S\ref{sec:closure-nonexistence} is easy to miss, because the natural way to fold a $\kapp$ correction into the energy budget is concrete, mechanical, and wrong --- and it is the step anyone equipped with the standard machinery would take.

Within the standard framework, a $\kapp$ correction to the diagnosed temperature has an immediate consequence. The conductive loss is closed with $q_{\mathrm{SH}} = -\kappa_{0}\,T^{5/2}\,\nabla T$ evaluated at the EUV-diagnosed temperature. If that temperature is the tail-weighted $\Teff$ while the bulk sits at $\Tcore = (\kapp - 3/2)/\kapp \cdot \Teff$, the conduction term has been evaluated at the wrong temperature, and the correction looks mechanical: integrating $q_{\mathrm{SH}} \propto T^{5/2}\,\nabla T$ along the loop, substituting $\Tcore$ for $\Teff$ multiplies the conductive loss by $(\Tcore/\Teff)^{7/2} = (\kapp/(\kapp - 3/2))^{-7/2}$, a factor $\approx 1/25$ at $\kapp = 2.5$. The \citet{withbroe1977} fluid budget falls by $\sim 47\%$ (relative to their reported coronal total of $3 \times 10^{5}$ erg cm$^{-2}$ s$^{-1}$); the long-standing excess of the inferred conductive loss over measured Alfv\'en-wave input narrows to within the measurement spread; and \citet{edmonds2026b}~\S 4.2.5 obtains the same $3$--$25\times$ overestimate from the identical substitution in the tokamak scrape-off layer. Every step is individually valid, the arithmetic is correct, and the result is the kind of concrete, falsifiable number a budget calculation exists to produce.

It is also meaningless, and \S\ref{sec:closure-nonexistence} is the reason: the factor $(\Tcore/\Teff)^{7/2}$ corrects the temperature in a coefficient that does not exist. There is no finite local $\kappa_{0}$ at $\kapp = 2.5$ for any temperature to be the argument of---and $\Tcore$ itself is a core-Maxwellian fit \citep{oka2013}, a description of the bulk, not a transport moment of the conducting tail. The trap is sprung most sharply by what the closed form returns when its poles are ignored and it is evaluated at $\kapp = 2.5$: a finite $-16.5\,\lambda_{\mathrm{M}}$---finite, concrete, and negative (Fig.~\ref{fig:closure}b). This is an analytic continuation of a divergent integral; it corresponds to no physical conductivity. The $47\%$ reduction is the same artifact one step removed, a ratio of two such evaluations. The very concreteness that makes the correction persuasive is the signature of a formula evaluated outside its domain: the value extracted depends on where the divergent integral is truncated (here $\propto v_{c}^{3}$ in the cutoff speed $v_{c}$), not on the plasma.

The correction is not a careless mistake; it is the natural move, and its invisibility is structural. The premise it rests on---that conduction is a local process characterized by a single temperature---is not an assumption one would think to check, because it is the constitutive assumption of fluid transport itself: it is the language in which the budget is written. The question ``which temperature should Spitzer-H\"arm use?'' is therefore malformed at a level above any step in the calculation; it presupposes that a right temperature exists, which presupposes the closure applies. This is the same circular structure identified on the inference side by \citet{edmonds2026b}: a Maxwellian tool operated to produce an answer that presupposes the tool's validity---there, an ionization-gated diagnostic returning $\Teff$ regardless of the distribution; here, a closure accepting a temperature regardless of whether a closure exists. The degeneracy result of \S\ref{sec:results} makes the closure-side version inescapable: the standard diagnostics cannot register the non-Maxwellianity that invalidates the closure, so the trap is not self-correcting within standard practice. The calculation is the natural one, and any $\kapp$-corrected fluid energy budget invites the same trap.

The distance from validity is quantitative. The relative entropy of the $\kapp = 2.5$ distribution from its energy-matched Maxwellian projection is $D_{\mathrm{KL}} = 0.32$ nats, derived in closed form in Appendix~\ref{app:dkl}; by the entropy-deficit identity \citep{jaynes1957,csiszar1975,levermore1996} this is exactly the information the Maxwellian closure discards. It is order unity---a still distinctly non-Maxwellian $\kapp = 6$ plasma sits a decade lower at 0.033 nats (Table~\ref{tab:dkl})---and it is model-independent: a definitional projection on the moments of the assumed distribution, involving no collision operator, atomic data, or regularization scheme. It measures how far the Maxwellian closure is from applicable at the observed $\kapp$, and the answer is: far.

\subsubsection{The shift, not the closure}\label{sec:awcomparison}

The standard quiet-Sun budget has been read as posing a sufficiency question for Alfv\'en-wave heating: the inferred coronal energy loss is dominated by the conductive term, $F_{c} \approx 2 \times 10^{5}$ erg cm$^{-2}$ s$^{-1}$ \citep[][Table 1]{withbroe1977}, against a measured wave flux at the coronal base that the Alfv\'en-wave community places in the range $0.5$--$5 \times 10^{5}$ erg cm$^{-2}$ s$^{-1}$ \citep{mcintosh2011,hahnsavin2014,soler2019,morton2025}, with a polar-coronal-hole outlier above \citep{hahnsavin2013} and order-of-magnitude downward corrections below \citep{tomczyk2007,goossens2013}; the 2025 picture, if anything, revises the wave-sufficiency framing downward \citep{morton2025apjl}. The gap between this conductive term and the wave flux has anchored part of the empirical case for impulsive heating in the quiet Sun.

That argument is not well-posed in its fluid-conductive form, and the present analysis cannot repair it---it can only say so. The conductive term $F_{c}$ on which the comparison rests is a Spitzer-H\"arm flux, and \S\ref{sec:closure-nonexistence} shows there is no Spitzer-H\"arm conductivity in this regime for $F_{c}$ to be the value of. Both the standard $F_{c}$ and any temperature-substituted ``correction'' to it are evaluations of a non-existent closure. What the analysis establishes is therefore a displacement of the question rather than a closed gap: the fluid-channel formulation of the Alfv\'en-wave-versus-conductive-budget argument loses its structural premise. The physical heat flux of a $\kapp \approx 2.5$ corona is real but non-local and tail-dominated (\S\ref{sec:closure-nonexistence}); the in-regime kinetic literature indicates it is enhanced over, and may reverse sign relative to, the Spitzer-H\"arm value rather than reduced \citep{landi2001,shoub1983,dorelliscudder2003}---opposite to the trap's apparent suppression, though we do not compute it here and the recasting in \S\ref{sec:closure-correction} does not depend on its direction. Whether the measured wave flux suffices is therefore a question about the total energy budget, including the non-local kinetic transport that no fluid closure represents---the open problem of \S\ref{sec:opentransport}, not a gap this paper closes.

\subsubsection{Open transport problems}\label{sec:opentransport}

Two transport-theory problems sit downstream of this calculation, preceded by a clarification that heads off a common misreading:

\paragraph{(a) The full kinetic heat flux for $\kapp \approx 2.5$.} \citet{landi2001} showed by direct kinetic simulation that at low $\kapp$ the conductive heat flux is enhanced over the Spitzer-H\"arm value and can reverse sign; \citet{shoub1983} and \citet{dorelliscudder1999,dorelliscudder2003} reach the same qualitative conclusion in the marginally collisional regime. As \S\ref{sec:closure-nonexistence} establishes, there is no local fluid conductivity to correct: the standard-$\kapp$ closed forms diverge across the range \citep{du2013,husidic2021,guodu2019} and the only finite (regularized) form is cutoff-dependent and of order $10^{4}$ \citep{husidic2022}; the generalized fluid-moment framework of \citet{cranmerschiff2021} likewise yields no closed-form $\kapp$-conductivity at $\kapp \approx 2.5$. The true conductive flux is therefore a fully kinetic quantity, not a Spitzer-H\"arm coefficient evaluated at any temperature. Computing it for $\kapp \approx 2.5$ under realistic coronal boundary conditions is an open problem, and no conclusion here depends on its magnitude or sign.

\paragraph{(b) Kinetic vs.\ fluid heat transport.} Direct measurements at solar-wind distances \citep{bale2013,pilipp1987,halekas2021} show the suprathermal-tail population (strahl and halo) carrying dominant heat flux non-locally. That measurement is real coronal physics; it is also not what Spitzer-H\"arm has been computing in the standard quiet-Sun budget. Spitzer-H\"arm is a moment closure valid only when the distribution sits within the Maxwellian family. Tail electrons escaping locally and depositing energy elsewhere along the field are described by kinetic transport equations, not by any moment closure of the local fluid hierarchy. The standard energy-budget calculation has been computing a fluid-bulk number from a closure that, at $\kapp \approx 2.5$, does not exist (\S\ref{sec:closure-nonexistence}); the kinetic strahl/halo flux has been outside that calculation entirely and is the open problem of (a). \citet{edmonds2026b} \S 6.1 engages the same distinction in the SOL context. As an upper bound for that open problem: under the \citet{shoub1983} $v^{4}$ collision-frequency scaling at QS bulk Knudsen number $\sim 0.05$, the velocity threshold above which electrons become collisionless over the relevant gradient scale is $v_{\mathrm{crit}}/v_{\mathrm{th}} \approx 2.1$, corresponding to $E \gtrsim 4.5\, k\Teff$. Numerical integration of the $\kapp = 2.5$ distribution shows this collisionless tail comprises $\sim 5\%$ of the electrons but carries $\sim 40\%$ of the total electron kinetic energy: the energy reservoir available to non-local transport. The heat flux itself is a higher (third) velocity moment, weighted by $v^{3}$ still further toward the tail, and is not computed here. The ceiling is generous for a second reason: the self-consistent ambipolar electrostatic potential that maintains quasi-neutrality against preferential electron escape throttles the collisionless tail, reshaping both $F_{k}$ and the transition-region energy balance. The same term inherits the structural role conduction plays in the standard paradigm: with no valid local fluid $F_{c}$ to perform it, the spatial redistribution that stabilizes static loop equilibria in conductively dominated models would have to be carried by $F_{k}$; whether it does so, and with what effective profile, is part of this open problem rather than a result of this paper.

\subsubsection{Implication for the nanoflare interpretation in the quiet Sun}\label{sec:nanoflarepillars}

The impulsive-heating interpretation of the corona has been developed in a broad literature \citep{klimchuk2006,klimchuk2015,reale2014}, with active-region-specific DEM-based constraints \citep{warren2012,cargill2014} providing the most quantitative tests of nanoflare frequency distributions. Of the QS-specific arguments, two bear directly on the AIA-imaging DEM-shape and energy-budget claims engaged in this paper:

\begin{itemize}
\item The apparent multi-thermal width of the quiet-Sun DEM with a hot tail at $\log T \approx 6.7$--6.8 has been read as evidence of multi-T plasma along every line of sight. Under the convergence-principle degeneracy verified in \S\ref{sec:results}, the AIA-imaging pipeline is one more diagnostic that returns $\Teff$ regardless of source family; the recovered FWHM range that real QS observations produce is also produced by single-T $\kapp$ emission, by Brooks-shape multi-T Maxwellian sources, and by combinations of the two (\S\ref{sec:demshape}). The DEM-width pillar does not uniquely identify multi-T plasma.
\item The gap between Alfv\'en-wave flux measurements and the standard required-heating budget has been read as evidence the quiet Sun needs a heating channel beyond AW dissipation. That gap rests on a Spitzer-H\"arm conductive term that does not exist for a $\kapp \approx 2.5$ corona (\S\ref{sec:closure-nonexistence}), so its fluid-conductive formulation is ill-posed (\S\ref{sec:awcomparison}). Whether AW magnitude suffices against the total budget (which depends on non-local kinetic transport not captured by any fluid closure) is the same open question it was under the standard accounting.
\end{itemize}

Other pillars in the broader nanoflare case are AR-specific or orthogonal to the $\kapp$ alternative. Hi-C braiding \citep{cirtain2013} is an AR observation of dynamic loop substructure. Time-lag analysis \citep{viall2012} studies temporal coherence patterns largely in AR. IRIS Si IV blueshifts \citep{testa2014} target electron-beam-driven impulsive events in AR transition-region brightenings. AR-core hot-plasma detections \citep{ishikawa2017,hannah2016,brosius2014} target a regime physically distinct from the quiet Sun. None are softened or affected by the present analysis, which restricts itself to the QS regime.

Within that scope, the nanoflare interpretation may remain physically real. The two QS-specific empirical arguments engaged here --- AIA-imaging DEM shape and the standard-accounting fluid-conductive-budget gap --- no longer support the impulsive-heating reading in the form they have historically been deployed: the DEM shape because the AIA pipeline cannot uniquely identify it, and the fluid-conductive-budget gap because that gap rests on a Spitzer-H\"arm conductive term that does not exist for a $\kapp \approx 2.5$ corona (\S\ref{sec:closure-nonexistence}). The total energy required to balance coronal losses is not reduced by this analysis: the conductive loss is a non-local kinetic flux of the suprathermal tail (\S\ref{sec:opentransport}~(b)), not a Spitzer-H\"arm fluid term, so the total-budget AW question persists, now correctly posed. What changes is narrower than the impulsive-heating literature has read it: the \emph{fluid-channel formulation} of the AW-vs-heating-budget argument loses its structural assumption, not the underlying question of whether AW magnitude suffices against the total budget the heating mechanism must supply.

The framework is compatible with --- and possibly explained by --- impulsive reconnection. A $\kapp \approx 2.5$ distribution at QS coronal densities would Coulomb-thermalize toward a Maxwellian on the bulk-collision timescale $\tau_{ee} \sim T_{e}^{3/2}/(n_{e} \ln\Lambda) \sim$ few $\times 10^{-3}$ s at $n_{e} \sim 10^{9}$ cm$^{-3}$ and $T_{e} \sim 1$ MK. It is additionally depleted by non-local strahl/halo outflow (\S\ref{sec:opentransport}~(b)) and exospheric solar-wind escape. Persistence at the observed level therefore requires continuous replenishment of the suprathermal tail. Two facts keep the energetic cost of that requirement in proportion. First, electron--electron Coulomb collisions conserve the total kinetic energy of the electron population: tail--bulk exchange is an internal redistribution, not a loss, so maintaining the distribution's \emph{shape} imposes no net power demand beyond the system's actual losses --- the radiative, conductive, wind, and kinetic-escape terms of the coronal energy budget. Second, even the internal recycling is slower than the bulk collision time suggests: the Coulomb relaxation time scales as $v^{3}$, so the energetically dominant tail relaxes an order of magnitude more slowly than the core, and electrons above the \citet{shoub1983} threshold of \S\ref{sec:opentransport}~(b) are effectively collisionless over the relevant gradient scale. The net external requirement is therefore the budget itself, not an additional sink; quantifying its kinetic-escape component is the open $F_{k}$ problem. Magnetic reconnection at small scales is among the most efficient known mechanisms for generating non-thermal suprathermal-electron tails \citep{drake2006} and is a natural \emph{local} candidate for that replenishment; non-local supply --- suprathermal electrons arriving along the field from elsewhere, the inflow counterpart of the strahl/halo outflow of \S\ref{sec:opentransport}~(b) --- is the complementary and equally open pathway, and the present analysis does not discriminate between them. In either case, the physics that generates the $\kapp$ structure is the physics that maintains it.

Under that reading the $\kapp \approx 2.5$ distribution may be the kinetic-distribution-level signature of nanoflare heating rather than evidence against it: the empirical case is reframed, not contested. AIA-imaging DEM analysis and the standard fluid-conductive-budget gap stop functioning as discriminators between $\kapp$-driven and Maxwellian-multi-thermal pictures of the corona, while neither is evidence against nanoflares as the underlying mechanism that maintains the $\kapp$ distribution against Coulomb relaxation.

%------------------------------------------------------------------
\subsection{What this leaves open, and for whom}\label{sec:openleft}
%------------------------------------------------------------------

Two claims stand: standard AIA-imaging DEM analysis cannot distinguish a single $\kapp$ distribution from a multi-thermal Maxwellian plasma in the quiet Sun (\S\ref{sec:results}); under that hypothesis, the Spitzer-H\"arm conductive closure underlying the standard quiet-Sun fluid energy budget does not exist (\S\ref{sec:closure}), so the fluid-conductive formulation of the Alfv\'en-wave-sufficiency argument is ill-posed and the budget question shifts to the non-local kinetic transport that no fluid closure represents.

\paragraph{Falsifiable predictions of the $\kapp$ framework (\citealt{edmonds2026a} \S 8.4).} The $\kapp$ premise used in this paper carries four direct observational tests independent of AIA-imaging DEM analysis:

\begin{enumerate}
\item \emph{AR core collapse.} Radio brightness measurements of high-density active-region cores at meter wavelengths (NRH, LOFAR, MWA) should show $T_{B}/T_{H} \lesssim 1.5$ as collisionality restores thermal equilibrium, in contrast to the quiet Sun's $\sim 2.4$.
\item \emph{Density dependence.} Multi-diagnostic campaigns combining radio brightness and spectroscopic density profiles should show the diagnostic ratio decreasing systematically with electron density, tracing the transition from tail-dominated to collision-dominated regimes.
\item \emph{Topological control.} At matched electron density, closed-field regions should show larger $R$ than open-field regions, reflecting the trapping of suprathermal electrons in closed-field topologies.
\item \emph{Forbidden-line detection.} Systematic quiet-Sun observations of the Fe X 6378 \AA{} forbidden line should reveal intensity enhancements consistent with low $\kapp$ \citep{dudik2014b}.
\end{enumerate}

\paragraph{Structural predictions specific to this paper's AIA-imaging analysis.}

\begin{enumerate}\setcounter{enumi}{4}
\item \emph{Fe XI charge-state crossover at non-QS effective temperatures.} The crossover ion at which tail-driven ionization and bulk-driven recombination balance is determined by the ratio of ionization potential to thermal energy. At $\Teff = 1.5$ MK the crossover lands at Fe XI (\S\ref{sec:fexi}). At a different $\Teff$ the crossover shifts to a different ion; the prediction is internally testable by independent atomic-physics evaluations against \citet{dzifcakova2023} ion-fraction tables across the coronal temperature range.
\item \emph{EUV continuum reversal at AIA wavelengths.} Free-free continuum at $\kapp = 2.5$ is suppressed near the thermal energy ($E \approx k\Teff$, 94 \AA) and enhanced at lower photon energies (171--335 \AA), with crossover at 131 \AA{} (\S\ref{sec:euvreversal}). Any analysis that models $\kapp$ continuum in the EUV should account for this energy-dependent behavior; coronagraphic continuum measurements that isolate free-free emission from line contributions provide a direct test.
\end{enumerate}

\paragraph{Compute follow-ups.}

\begin{enumerate}\setcounter{enumi}{6}
\item \emph{$\kapp$-modified excitation rates for visible forbidden lines.} Fe X 6378 \AA{} (the red coronal line) is identified by \citet{dudik2014b} as the most conspicuous exception to the general $\kapp$-insensitivity of EUV/visible line ratios within Fe IX--XIII, with intensity enhancement up to a factor of two at low $\kapp$; a calculation using KAPPA-native collision strengths paired with spatially resolved quiet-Sun observations would provide a direct test complementary to the DEM-level analysis.
\item \emph{SphinX soft X-ray spectrum.} At $E \gg k\Teff$ the $\kapp$ continuum enhancement is dramatic. The SphinX quiet-Sun spectrum \citep{sylwester2019}, which requires a two-component Maxwellian fit, may be consistent with a single-$\kapp$ bremsstrahlung spectrum.
\end{enumerate}

\paragraph{Open transport-theory problems.}

\begin{enumerate}\setcounter{enumi}{8}
\item \emph{Full kinetic closure for $\kapp \approx 2.5$ at $\mathrm{Kn} = 0.01$--$0.1$.} \citet{landi2001} showed flux reversal at low $\kapp$; the quantitative impact on the QS energy budget under realistic boundary conditions is unsolved. Partial generalized $\kapp$-conductivity derivations \citep{dorelliscudder1999,dorelliscudder2003,du2013,cranmerschiff2021} sketch the territory; as \S\ref{sec:closure-nonexistence} shows, there is no fluid component to estimate --- the local closure does not exist, and the true conductive flux is the fully kinetic quantity this problem must compute.
\item \emph{Dissipation pathway for the measured AW flux.} The energy is available; the mechanism is the next question. Wave-particle resonance pathways that dissipate AW flux are natural candidates for whatever mechanism maintains the $\kapp$ distribution at $\kapp \approx 2.5$ against Coulomb relaxation; impulsive reconnection is a second candidate.
\end{enumerate}

%==================================================================
\FloatBarrier

\section{Conclusions}\label{sec:conclusions}
%==================================================================

A single isothermal $\kapp$-distributed plasma at $\Tcore = 0.6$ MK, $\Teff = 1.5$ MK, $\kapp = 2.5$ (the central value from \citet{edmonds2026a}; a pipeline probe, not a model of the corona) was processed through the standard SDO/AIA + \citet{hannah2012} regularized DEM inversion pipeline alongside two multi-thermal source families.

The pipeline returned a multi-thermal Maxwellian DEM with $\chisq/\dof = 1.00$ at FWHM $= 0.222$ in $\log T$ for the single-T probe, sitting just below the narrow edge of the FWHM distribution the same pipeline returns from 80 real quiet-Sun AIA patches at solar minimum (median 0.283, range 0.230--0.401, stable across two minimum dates). The two multi-thermal synthetic sources --- Brooks-shape Maxwellian and multi-T $\kapp = 2.5$ --- recover FWHMs of 0.319 and 0.305, both inside the real-QS distribution. The pipeline did not distinguish any of the three source families from genuine multi-thermal structure, as predicted by the convergence principle: every ionization-gated diagnostic structurally returns $\Teff$ regardless of source family. Hinode/EIS within-ion line ratios break the AIA-imaging degeneracy \citep{dudik2014b,dudik2019}, complementary to \citet{lorincik2020}'s Maxwellian-consistent QS EIS finding. Two structural features specific to the solar implementation emerged: the iron charge-state crossover at Fe XI, where tail-driven ionization and bulk-driven recombination balance; and a free-free continuum reversal at AIA wavelengths, suppressed near $E \approx k\Teff$ and enhanced below. Absolute-radiance self-consistency holds against \citet{brooks2009} line radiances, with a 1.06 dex single-T tilt across Fe IX--XVI that collapses to 0.14 dex under a multi-T $\kapp$ source.

The consequence runs further than the diagnostic. The Spitzer-H\"arm conductivity used in the standard quiet-Sun energy budget is a Maxwellian moment closure evaluated at the EUV-derived $\Teff$, a tail-weighted moment of a non-Maxwellian distribution. The substitution of $\Tcore$ for $\Teff$ that the standard budget invites yields a concrete conductive-term reduction, but it corrects a coefficient that does not exist: the local Spitzer-H\"arm conductivity has no finite positive value across the entire $\kapp \in [2,3]$ range --- the conductivity integral diverges, and the finite value the formula returns at $\kapp = 2.5$ is an analytic continuation of that divergent integral (\S\ref{sec:closure-nonexistence}). The non-existence is a property of the distribution class, not of the Sun: it holds for any plasma whose electrons sit at $\kapp \in [2,3]$, with the quiet corona as the worked example. The standard fluid-conductive energy budget thus rests on a closure invalid for a $\kapp \approx 2.5$ corona; \citet{edmonds2026b} \S 4.2.5 reaches the same conclusion in the tokamak SOL context. The contribution is a shift of the question, not a closure of the gap: the true conductive flux is non-local and fully kinetic, outside any fluid moment closure, and whether the measured Alfv\'en-wave flux ($0.5$--$5 \times 10^{5}$ erg cm$^{-2}$ s$^{-1}$) suffices is a question about that total kinetic budget --- the open problem flagged in \S\ref{sec:opentransport}.

Two QS-specific empirical pillars supporting impulsive-heating interpretations lose their structural assumptions: the multi-thermal DEM-width pillar under the convergence-principle degeneracy, and the AW-vs-fluid-conductive-budget pillar because that budget's conductive term rests on a closure that does not exist in this regime. Other pillars in the broader nanoflare case are AR-specific or orthogonal --- Hi-C braiding (AR loop substructure), time-lag analysis (largely AR), IRIS Si IV blueshifts (AR TR brightenings), AR-core hot-plasma detections \citep{ishikawa2017,hannah2016,brosius2014} --- and are not affected by the present analysis, which restricts itself to the QS regime. Two transport-theory problems sit downstream: full kinetic closure for $\kapp \approx 2.5$ at $\mathrm{Kn} = 0.01$--0.1, and the dissipation pathway for the measured AW flux into bulk and tail components. Both are problems the kinetic-theory and wave-physics communities are equipped to solve.

%==================================================================
\section*{Code and Data Availability}
%==================================================================

The analysis pipeline, sensitivity tests, and pre-computed numerical results
are publicly available at
\url{https://github.com/Final-Stop-Consulting/kappa-dem-inversion}, archived
on Zenodo with concept DOI
\href{https://doi.org/10.5281/zenodo.19188096}{10.5281/zenodo.19188096},
which resolves to the version-specific DOI of the release accompanying this
paper. The pipeline is implemented in Python and does not require IDL or
SolarSoft; computational dependencies are listed in the repository
\texttt{requirements.txt}. The Stage-2 per-ion contribution checkpoint is
committed to the repository so that the sensitivity scripts of
\S\ref{sec:noise}, \S\ref{sec:demshape}, and \S\ref{sec:emrec} can be
re-run independently of the main pipeline; a reviewer's verification map
in the repository \texttt{README.md} cross-references each numerical claim
in this paper to its corresponding script, result file, dependency tier,
and expected runtime. The $\kapp$-modified ion-fraction tables
\citep{dzifcakova2023} are distributed with the KAPPA package by the
Astronomical Institute of the Czech Academy of Sciences; the CHIANTI v11
atomic database \citep{dere2023,dufresne2024} is publicly available at
\url{https://www.chiantidatabase.org/}; the \citet{brooks2009} reference
quiet-Sun DEM is included with the CHIANTI v11.0.2 distribution as
\texttt{quiet\_sun\_eis.dem}.

%==================================================================
\section*{Software}
%==================================================================

This work made use of \texttt{ChiantiPy} \citep{dere2023}, \texttt{aiapy}
\citep{barnes2020}, \texttt{demregpy} \citep{hannah2012}, \texttt{numpy},
\texttt{scipy}, \texttt{matplotlib}, and \texttt{astropy}.

%==================================================================
% References
%==================================================================

\FloatBarrier

\bibliographystyle{plainnat}

\appendix
%==================================================================
\section{Relative entropy of the kappa distribution from its Maxwellian projection}\label{app:dkl}
%==================================================================

This appendix derives the closure-validity measure used in \S\ref{sec:closure-correction}: the relative entropy (Kullback--Leibler divergence) of the $\kapp$ velocity distribution from its energy-matched Maxwellian projection. The quantity is a definitional projection on the moments of the distribution --- no collision operator, atomic data, or inversion machinery enters --- and, being a divergence, it is invariant under coordinate and scale changes, so its value in nats is convention-free.

Normalize Eq.~(\ref{eq:kappa-vdf}) as a probability density. In terms of the scale parameter $w$ (with $v_{th}^{2} = (2\kapp - 3)\,w^{2}/\kapp$ recovering the convention of \S\ref{sec:kappa-ioneq}),
\begin{equation}
f_{\kapp}(\mathbf{v}) = N_{\kapp} \left[ 1 + \frac{v^{2}}{\kapp w^{2}} \right]^{-(\kapp+1)},
\qquad
N_{\kapp} = (\pi \kapp w^{2})^{-3/2}\, \frac{\Gamma(\kapp+1)}{\Gamma(\kapp-1/2)},
\qquad
\langle v^{2} \rangle = \frac{3 \kapp w^{2}}{2\kapp - 3}.
\label{eq:fkapp-norm}
\end{equation}
The energy-matched Maxwellian projection is $f_{\mathrm{M}}(\mathbf{v}) = (2\pi\sigma^{2})^{-3/2} \exp(-v^{2}/2\sigma^{2})$ with $3\sigma^{2} = \langle v^{2} \rangle$. Matching the second moment anchors the projection at the mean-energy temperature $\Teff$ (\S\ref{sec:kappa-ioneq}), not the bulk $\Tcore$: the deficit is the information the best Maxwellian \emph{at the observed effective temperature} discards, being simultaneously too hot for the $\kapp$ core and too thin in its tail. Because $\ln f_{\mathrm{M}}$ is linear in the matched invariants $\{1, v^{2}\}$, the cross term in the divergence collapses, $\int f_{\kapp} \ln f_{\mathrm{M}} = \int f_{\mathrm{M}} \ln f_{\mathrm{M}}$, and the divergence reduces to the entropy deficit \citep{jaynes1957,csiszar1975,levermore1996}:
\begin{equation}
D_{\mathrm{KL}}(f_{\kapp} \,\|\, f_{\mathrm{M}}) = H(f_{\mathrm{M}}) - H(f_{\kapp}),
\label{eq:deficit}
\end{equation}
with $H$ the Gibbs--Shannon entropy. The Maxwellian entropy is $H(f_{\mathrm{M}}) = \tfrac{3}{2}\ln(2\pi e \sigma^{2})$. For the $\kapp$ distribution, $H(f_{\kapp}) = -\ln N_{\kapp} + (\kapp+1)\,\mathrm{E}[\ln(1 + v^{2}/\kapp w^{2})]$, and the expectation follows by differentiating the normalization integral with respect to the exponent (equivalently, from the trivariate Student-$t$ correspondence with $\nu = 2\kapp - 1$ degrees of freedom): $\mathrm{E}[\ln(1 + v^{2}/\kapp w^{2})] = \psi(\kapp+1) - \psi(\kapp-1/2)$, with $\psi$ the digamma function. The scale $w$ cancels, leaving a function of $\kapp$ alone:
\begin{equation}
D_{\mathrm{KL}}(\kapp) = \frac{3}{2} \ln\!\left[ \frac{2e}{2\kapp - 3} \right]
+ \ln\frac{\Gamma(\kapp+1)}{\Gamma(\kapp-1/2)}
- (\kapp+1)\left[ \psi(\kapp+1) - \psi(\kapp-1/2) \right].
\label{eq:dklclosed}
\end{equation}
Equation~(\ref{eq:dklclosed}) vanishes as $\kapp \to \infty$ (the Maxwellian limit) and diverges as $\kapp \to 3/2$. Table~\ref{tab:dkl} evaluates it on the quiet-Sun $\kapp$ range and at a sub-critical reference value; the closed form agrees with direct numerical quadrature of the defining integral to machine precision.

\begin{table}[!htbp]
\centering
\caption{Information discarded by the Maxwellian closure: $D_{\mathrm{KL}}$ of the $\kapp$ distribution from its energy-matched Maxwellian projection, Eq.~(\ref{eq:dklclosed}).}
\label{tab:dkl}

\begin{tabular}{lrrrr}
\toprule
$\kapp$ & 2 & 2.5 & 3 & 6 \\
\midrule
$D_{\mathrm{KL}}$ (nats) & 0.695 & 0.320 & 0.187 & 0.033 \\
\bottomrule
\end{tabular}

\end{table}

Across the quiet-Sun range $\kapp \in [2,3]$ the deficit is 0.19--0.69 nats --- order unity, with 0.32 at the central value --- an order of magnitude above the $\kapp = 6$ reference point, which is itself still distinctly non-Maxwellian. This is the quantitative sense in which the Spitzer-H\"arm closure is at the threshold of its validity for the observed plasma (\S\ref{sec:closure-correction}).

\end{document}